%% file: arxiv_main.tex
\documentclass[ms]{informs3}

\allowdisplaybreaks
\RequirePackage{tgtermes}
\RequirePackage{newtxtext}
\RequirePackage{newtxmath}
\RequirePackage{bm}
\RequirePackage{endnotes}
\usepackage{natbib}

\usepackage{algorithm}
\usepackage[noend]{algpseudocode}

\usepackage{float}
\bibpunct[, ]{(}{)}{,}{a}{}{,}
\def\bibfont{\small}

\usepackage{nicefrac}

\usepackage{color-edits}
\addauthor{srs}{orange}
\addauthor{rs}{teal}
\addauthor{mu}{indigoblue}

\usepackage{subcaption}

\usepackage{mathtools}
\usepackage{booktabs}
\usepackage{subcaption}
\usepackage{adjustbox}
\usepackage{makecell}
\usepackage{xspace}
\usepackage{comment}
\usepackage[shortlabels]{enumitem}
\usepackage{tikz}
\usepackage{pgfplots}
\pgfplotsset{compat=1.18}
\usetikzlibrary{patterns,positioning,angles,quotes,calc,arrows.meta,decorations.pathreplacing}
\usepgfplotslibrary{fillbetween}
\usepackage[colorlinks=true,breaklinks=true,bookmarks=true,urlcolor=magenta,citecolor=magenta,linkcolor=magenta,bookmarksopen=false]{hyperref}
\usepackage[capitalize,noabbrev]{cleveref}



\newcommand{\IC}{\textsf{IC}}
\newcommand{\bfx}{\mathbf{x}}
\newcommand{\bfalpha}{\boldsymbol{\alpha}}

\newcommand{\E}{\mathbb{E}}

\ifdefined\argmin\else\DeclareMathOperator*{\argmin}{arg\,min}\fi
\ifdefined\argmax\else\DeclareMathOperator*{\argmax}{arg\,max}\fi

\newcommand{\X}{\mathcal{X}}
\newcommand{\cl}{\operatorname{cl}}
\newcommand{\abs}[1]{\left|#1\right|}
\newcommand{\T}{\mathcal{T}}

\newcommand{\DeltaN}{\Delta^{\mathrm{N}}}

\newtheorem{observation}{Observation}

\definecolor{royalblue}{RGB}{54,97,223}
\definecolor{nupurple}{RGB}{64,31,104}
\definecolor{kulblue}{RGB}{36,146,178}
\definecolor{darkorange}{RGB}{205,144,50}
\tikzset{
  paper figure text/.style={font=\scriptsize},
  paper tick label/.style={font=\scriptsize},
  paper axis/.style={draw=black!80,line width=0.75pt},
  paper grid/.style={draw=black!18,densely dotted,line width=0.55pt},
  paper benchmark/.style={draw=black!55,densely dashed,line width=0.80pt},
  paper primary/.style={draw=royalblue,line width=1.15pt},
  paper auxiliary/.style={draw=black!45,line width=0.75pt}
}
\pgfplotsset{
  paper plot/.style={
    width=0.94\linewidth,
    height=0.60\linewidth,
    xmin=0.1,
    xmax=2.1,
    xtick={0.2,0.4,0.6,0.8,1.0,1.2,1.4,1.6,1.8,2.0},
    xlabel={Tightness $t$},
    grid=major,
    axis background/.style={fill=white},
    label style={font=\scriptsize},
    tick label style={font=\scriptsize},
    legend style={draw=black!20,fill=white,fill opacity=0.92,text opacity=1,
      rounded corners=1pt,font=\scriptsize,cells={anchor=west},inner sep=2pt,
      at={(0.02,0.98)},anchor=north west},
    every axis plot/.append style={line join=round},
    clip=false,
  },
  paper series one/.style={royalblue,line width=1.15pt,mark=*,mark size=1.65pt},
  paper series two/.style={darkorange,line width=1.15pt,mark=square*,mark size=1.55pt},
  paper series three/.style={kulblue,line width=1.15pt,mark=triangle*,mark size=1.75pt},
  paper band one/.style={draw=none,fill=royalblue,fill opacity=0.13},
  paper band two/.style={draw=none,fill=darkorange,fill opacity=0.13},
  paper band three/.style={draw=none,fill=kulblue,fill opacity=0.13}
}
\OneAndAHalfSpacedXI
\EquationsNumberedThrough
\TheoremsNumberedThrough
\MANUSCRIPTNO{}

\renewenvironment{proof}[1][Proof]{%
  \par\noindent{\itshape #1.}\ \ignorespaces
}{%
  \hfill\Halmos\par
}

\begin{document}

\RUNAUTHOR{Salamat, Sinclair, and Iravani}
\RUNTITLE{Incentive Compatible Mechanisms for Sequential Fair Resource Allocation with Private Demands}
\TITLE{Incentive Compatible Mechanisms for Sequential Fair Resource Allocation with Private Demands}
\ARTICLEAUTHORS{%
\AUTHOR{Ramtin Salamat}
\AFF{Kellogg School of Management,
Northwestern University, \EMAIL{ramtin.salamat@kellogg.northwestern.edu}}

\AUTHOR{Sean R. Sinclair}
\AFF{Department of Industrial Engineering and Management Sciences,
Northwestern University, \EMAIL{sean.sinclair@northwestern.edu}}

\AUTHOR{Seyed M.R. Iravani}
\AFF{Department of Industrial Engineering and Management Sciences,
Northwestern University, \EMAIL{s-iravani@northwestern.edu}}
}

\ABSTRACT{%
\input{short_parts/abstract}}
\KEYWORDS{Sequential fair allocation; incentive compatibility; mechanism design without transfers; dynamic programming}

\maketitle

\input{short_parts/introduction}
\input{short_parts/related_work}

\input{short_parts/preliminary}
\input{short_parts/structure_ic}
\input{short_parts/cap_optimality}
\input{short_parts/approximation}
\input{short_parts/simulations}
\input{short_parts/conclusion}

\clearpage
\bibliographystyle{informs2014}
\bibliography{sequential}

\ECSwitch
\input{short_parts/app_simulations}

\end{document}

%% file: short_parts/abstract.tex
We study a strategic variant of the canonical online fair allocation problem, where a planner must sequentially allocate a divisible resource to agents with privately known demands. At each stage, the planner announces an allocation rule, the agent reports their demand, and the planner allocates resources according to the rule and the reported demand. Unlike the standard model, we allow agents to strategically misreport their needs, as can arise in nonprofit settings where demand is self-reported and monetary transfers are unavailable.
We characterize the structure of incentive-compatible (IC) allocation rules in this model, and use this to identify conditions under which the optimal mechanism has a simple cap structure. We then develop an approximation method for the optimal IC mechanism through a careful reduction of the discretized dynamic program to a longest-path problem on a directed acyclic graph. Numerical experiments show that ignoring strategic reporting can be substantially more costly than imposing incentive compatibility.

%% file: short_parts/introduction.tex
\section{Introduction}
\label{sec:introduction}

Many operational systems allocate scarce resources sequentially to agents with uncertain and heterogeneous needs, including nonprofit distribution, humanitarian relief, and medical supply allocation~\citep{sengul2023equitable,GerdingPerezDiazAzizGaspersMarcuMatteiWalsh2019,ghodsiDominantResourceFairness2011a}. A central planner must decide how much of a limited resource to allocate to each arriving agent based on their reported demand, without knowing the demands of future agents, with the goal of allocating resources {\em fairly}~\citep{Varian1974,BertsimasFariasTrichakis2011,SinclairJainBanerjeeYu2023}. A central difficulty is that the planner may need to rely on agents' self-reported needs, particularly in nonprofit settings where demand is difficult to verify and monetary transfers are unavailable. For example, during the COVID-19 pandemic, the nonprofit GetUsPPE received more than 7,000 PPE requests within three weeks~\citep{Schlanger2020,BalaLeePallantEtAl2021}. The organization allocated supplies using healthcare facilities' self-reported needs.

However, self-reported demand creates an incentive problem since agents may overstate their need to obtain more resources, specifically when the resource is limited.  During the COVID-19 pandemic, the U.S. Government Accountability Office reported concerns that states may have requested more supplies than needed and that unreliable demand information impeded allocation to areas of greatest need~\citep{USGAO2021}. At the same time, these settings place substantial importance on equitable allocation; both Feeding America and the World Health Organization emphasize distributing scarce resources according to need~\citep{FeedingAmerica2022,WHO2020ResourceAllocation}.

The sequential fair resource allocation problem has received substantial attention in the operations literature, with recent work studying the structure of optimal allocation policies under truthful agents~\citep{ManshadiNiazadehRodilitz2023,gao2026sequentialfairallocationrouting}. However, less is known about how strategic reporting changes the problem. We therefore ask: 
\begin{center}
{\em What do incentive-compatible sequential allocation rules look like, and how can we solve for or approximate the optimal incentive-compatible mechanism for maximizing fairness?}
\end{center}

\subsection{Main Contributions}

We study fair sequential allocation, building on \citet{LienIravaniSmilowitz2014} by assuming agents have private demands and single-peaked utility around their true demand: allocations below demand increase utility, while allocations above demand incur an overage penalty. Our main contributions are:
\begin{itemize}
    \item {\bf Characterization of incentive-compatible (IC) rules}. We characterize all deterministic incentive-compatible rules through the set of allocation levels the planner makes attainable. Each reported demand is assigned the attainable allocation that the agent prefers, accounting for the different costs of receiving too little or too much. Importantly, this characterization is independent of the planner's choice of objective, and therefore applies broadly across different notions of fairness (\cref{thm:IC-structure}).
 
    \item \textbf{Optimality of cap mechanisms.}
    We then show that for our recursive max-min fairness objective, when over-allocation is sufficiently inexpensive, an optimal incentive-compatible rule takes the simple form of a cap mechanism, allocating exactly the reported demand up to a pre-specified cap (\cref{thm:two_stage_cap}). Because a cap mechanism is characterized by a single parameter, this reduces the planner's optimization problem to choosing the optimal cap. However, we also show by counterexample that cap mechanisms need not be optimal when over-allocation is sufficiently costly.

    \item \textbf{Tractable approximation.}
    When cap mechanisms are not optimal, finding the optimal IC rule requires optimizing over a complex class of attainable allocation sets. Using our characterization of IC rules, we discretize the allocation space and show that the resulting optimization problem at each stage can be represented as a longest-path problem on a directed acyclic graph. This avoids enumerating all possible attainable sets and yields a polynomial-time algorithm for the discretized problem. We further show that the resulting value converges to the optimal continuous value as the grid is refined, providing an additive approximation guarantee (\cref{prop:disc_complexity,thm:grid_approx}).

    \item \textbf{Impact of strategic reporting.} 
    Lastly, we use numerical simulations to show that the loss from imposing incentive compatibility is relatively modest compared to the loss that can arise when a planner ignores strategic reporting (\cref{sec:numerics}). We observe that ignoring strategic reporting is most costly at intermediate resource levels, and the gains from incentive-compatible design are largest, on average, when demand variability is low and the overage penalty is high.
\end{itemize}
Taken together, these results show that strategic reporting fundamentally changes the sequential fair allocation problem from~\citet{LienIravaniSmilowitz2014,ma2023fairness,ManshadiNiazadehRodilitz2023,gao2026sequentialfairallocationrouting}, but that accounting for incentives need not make the problem intractable. It also highlights incentive design as an important consideration in many practical settings where demand is private information and therefore allocation relies on self-reported demands.

%% file: short_parts/related_work.tex
\subsection{Related Work}

Fair resource allocation has been widely studied in economics, computer science, and operations; see also \citet{AleksandrovWalsh2020} for a survey of online fair division.
\paragraph{Dynamic fair allocation without strategic demand.}
The most closely related work studies sequential fair allocation when agents truthfully report their demands~\citep{LienIravaniSmilowitz2014,ma2023fairness,ManshadiNiazadehRodilitz2023,liang2022efficiency, ma2025forward}. This literature characterizes optimal or near-optimal allocation policies under various notions of fairness, including ex-post and ex-ante minimum fill rates, and fairness-efficiency trade-offs.  More recently, \citet{gao2026sequentialfairallocationrouting} jointly optimize allocation and visitation order under ex-post and forward max-min fairness objectives. In contrast, agents in our model privately observe their demands and may strategically misreport them, so the planner must induce truthful reporting through the allocation rule itself.

\citet{BalcikIravaniSmilowitz2014} and \citet{AlkaabnehShehadehDiabat2023} incorporate routing and delivery decisions into nonprofit resource allocation, while \citet{DonahueKleinberg2020,ma2023fairness,SinclairJainBanerjeeYu2023,GerdingPerezDiazAzizGaspersMarcuMatteiWalsh2019} study fairness-efficiency trade-offs in dynamic, online, or sequential allocation environments. Related work in inventory routing and humanitarian logistics studies how current delivery and allocation decisions affect future resource availability; see, for example, \citet{BurnsHallBlumenfeldDaganzo1985,BassokErnst1995,BermanLarson2001,CoelhoCordeauLaporte2014,EisenhandlerTzur2019}.

\paragraph{Static fair allocation with incentive compatibility when monetary transfers are unavailable.}
A second related strand studies strategy-proof allocation without monetary transfers under single-peaked preferences. \citet{Sprumont1991} characterizes the uniform rule for dividing a fixed quantity of a divisible resource, while \citet{BarberaJacksonNeme1997,KlausPetersStorcken1998} study rules allowing asymmetric treatment or individual endowments. These papers study simultaneous allocation, whereas agents in our model arrive sequentially and each allocation reduces the remaining budget. Related work considers course allocation with indivisible objects~\citep{Budish2011}, multiple divisible goods without monetary transfers~\citep{ColeGkatzelisGoel2013}, and public-school assignment with heterogeneous services~\citep{AshlagiShi2016}.

\paragraph{Dynamic allocation with private information.}
A third related strand studies allocation with private information. \citet{KimKimIravaniPark2021} study sequential capacity allocation with private demand, \citet{ZhangAtasuAyerToktay2020} study truthful mechanisms for medical surplus allocation, and \citet{FanChenTang2023} study scarce-resource allocation with heterogeneous favoritism. \citet{blanchard2026near} use promised utilities to construct dynamic IC mechanisms approaching first-best welfare, and \citet{ZengBhattKoppelGanesh2025} develop a meta-algorithm that converts one-shot mechanisms into sequential ones while approximately preserving incentives and welfare. \citet{Prendergast2022} shows the value of local information about heterogeneous needs in food-bank allocation. Our paper differs by studying nonprofit allocation with private scalar demands and a minimum-fill-rate fairness objective, without monetary transfers. We characterize the resulting IC allocation rules and embed them in the planner's sequential dynamic program.

%% file: short_parts/preliminary.tex
\section{Preliminaries}
\label{sec:prelim}

\paragraph{Model primitives.}
We consider a sequential resource allocation problem with total budget \(B_1>0\) of a divisible resource. A decision-maker (also referred to as the ``planner'') sequentially visits $n$ agents indexed by $i \in [n]$. Each agent \(i\) has demand \(d_i\), drawn independently from a known distribution \(F_i\) with continuous density \(f_i\). Without loss of generality, we assume that \(d_i \in [0,1]\). We further assume $f_i(d)>0$ for all $d\in(0,1)$. The planner knows the distributions \((F_i)_{i\in[n]}\), but each realized demand \(d_i\) is observed only by agent \(i\). Abusing notation, when the subscript $i$ is omitted, we use the corresponding variable to refer to a generic agent or stage.
At each stage $i \in [n]$, the following occurs:
\begin{itemize}
    \item The planner visits agent $i$ and announces their allocation rule (also referred to as ``mechanism'') \(x_i(\cdot;B_i):[0,1]\to[0,B_i\wedge 1]\);
    \item Observing this rule, the agent reports a demand $\hat d_i \in [0,1]$ (we model reporting behavior shortly);
    \item The planner makes an irrevocable allocation $x_i(\hat d_i;B_i)\in[0,B_i\wedge 1]$ based on their allocation rule;
    \item The remaining budget is updated as $B_{i+1} = B_i - x_i(\hat{d}_i; B_i)$.
\end{itemize}
Throughout, we consider deterministic mechanisms and assume the planner commits to an allocation rule before observing the agent's reported demand~\citep{myerson1981optimal}.

\paragraph{Agent utility.}
We assume agents have single-peaked preferences over allocations \citep{Sprumont1991}. Each agent $i$ has a known parameter $\alpha_i > 0$ capturing their disutility from over-allocation beyond their {\em true} latent demand $d_i$, reflecting that excess resources may be wasteful or operationally costly \citep{sengul2023equitable}. Given demand \(d\), allocation \(x\), and overage penalty \(\alpha\), we define the utility function:
\begin{equation}
\label{eq:utility}
u(x,d;\alpha)=\min\{x,d\}-\alpha\max\{x-d,0\}.
\end{equation}

\paragraph{Reported demand and incentive compatibility.}
We assume agents are rational and report their demand to maximize their utility. Hence, agent $i$ chooses a report $\hat d_i$ based on their true demand $d_i$ that maximizes:
\[
\hat d_i \in \argmax_{\tilde d\in[0,1]}
u(x_i(\tilde d;B_i),d_i;\alpha_i).
\]
With this in place, we now turn to incentive-compatible mechanisms, that is, mechanisms under which truthful reporting is optimal for the agents so that $\hat d_i = d_i$.

\begin{definition}
For a generic stage with remaining budget \(B\) and overage penalty \(\alpha\), an allocation rule $x(\cdot;B):[0,1]\to[0,B\wedge 1]$ is {\bf incentive compatible} if
\[
u(x(d;B),d;\alpha) \geq u(x(\hat d;B),d;\alpha)
\qquad
\forall d,\hat d\in[0,1].
\]
Let \(\IC(B;\alpha)\) denote the set of all such allocation rules.
Given \(\boldsymbol{\alpha}=(\alpha_1,\ldots,\alpha_n)\), a dynamic mechanism \(\bfx=(x_1,\ldots,x_n)\) is incentive compatible if \(x_i(\cdot;B_i)\in \IC(B_i;\alpha_i)\) for every \(i\in[n] \) and every realized budget $B_i$. We denote the set of incentive-compatible dynamic mechanisms by \(\IC(\boldsymbol{\alpha})\).
\end{definition}

\paragraph{Planner's objective.}

We assume that the planner measures the utility for agents by their fill rate, the fraction of demand satisfied. If an agent with true demand $d \geq 0$ receives allocation $x$, then their {\em fill rate} is $\beta(x,d) = \min \left\{\frac{x}{d},1\right\}$, with $\beta(x,0) := 1$. Fill-rates are widely used in practice as a performance measure for resource allocation, including by organizations such as Feeding America~\citep{sengul2023equitable}.  Given a dynamic mechanism $\bfx=(x_1,\dots,x_n)$, define the realized fill rate at stage $i$ as \( \beta_i := \beta(x_i(d_i;B_i), d_i)\). (Note here that $\hat d_i = d_i$ since we restrict to IC mechanisms.) The planner's objective is then to maximize $V^{\bfx}(B_1)$, which is the agents' sequential ex-ante expected minimum fill-rate, defined as
\begin{equation}
\label{eq:ex_ante_fill_rate}
V^{\bfx}(B_1) := \E_{F_1}\left[ \beta_1 \wedge \E_{F_2}\left[ \ldots \E_{F_n}[\beta_n] \right] \right].
\end{equation}
This objective balances service across agents over time by comparing the current agent's fill rate with expected continuation performance. This matches the intuition that allocating more inventory now can improve current service but leaves less inventory for future agents.

We further note that the optimal value functions are equivalently defined recursively by
\begin{align}
\label{eq:optimal_recursion}
V_n(B) &:= \sup_{x \in \IC(B;\alpha_n)} \E_{d_n \sim F_n} \left[ \beta(x(d_n; B), d_n) \right],\\
V_i(B) &:= \sup_{x \in \IC(B;\alpha_i)} \E_{d_i \sim F_i}\left[ \beta(x(d_i; B), d_i) \wedge V_{i+1}(B - x(d_i; B)) \right], \qquad i = n - 1,\ldots,1. \label{eq:optimal_recursion_nonterminal}
\end{align}
Thus, the planner's optimal objective value is $V_1(B_1) = \sup_{\bfx \in \IC(\bfalpha)} V^{\bfx}(B_1)$.

%% file: short_parts/structure_ic.tex
\section{Characterizing Incentive-Compatible Mechanisms}
\label{sec:ic-characterization}

Our first result completely characterizes IC allocation rules. The characterization is independent of the planner's objective and depends only on agent utility through the overage penalty $\alpha$. In the following, we suppress the agent index $i$ and write $d$, $\alpha$, and $x(\cdot;B)$ for a generic agent.

\begin{theorem}
\label{thm:IC-structure}
For $B\ge0$ and $\alpha>0$, an allocation rule
$x(\cdot;B):[0,1]\to[0,B\wedge1]$ is incentive compatible if and only if there exists a nonempty closed set $\mathcal X \subseteq [0, B\wedge1]$ such that if $m := \inf \mathcal X$ and $C := \sup \mathcal X$, then $x$ is constructed as follows:
\begin{enumerate}
    \item \textbf{Bottom plateau.} For all $d < m$, $x(d;B) = m$.
    \item \textbf{Exact allocation on attainable points.} For all $d \in \mathcal X$, $x(d;B) = d$.
    \item \textbf{Jump blocks across gaps.} Since $(m,C) \setminus \mathcal X$ is open, it can be written as $(m,C) \setminus \mathcal X = \bigcup_j (v_j, u_{j+1}),$ where the endpoints $(v_j, u_{j+1}) \in \mathcal X^2$. For each gap $(v_j, u_{j+1})$, define $\tau_j := \frac{v_j + \alpha u_{j+1}}{1 + \alpha}$. Then
    \[
    x(d;B)=v_j
    \quad \text{for } d\in (v_j,\tau_j),
    \qquad
    x(d;B)=u_{j+1}
    \quad \text{for } d\in (\tau_j,u_{j+1}),
    \]
    with arbitrary tie-breaking at $d=\tau_j$.
    \item \textbf{Top plateau.} For all $d > C$, $x(d;B) = C$.
\end{enumerate}
\end{theorem}

\begin{figure}[!t]
    \centering
    \resizebox{0.7\textwidth}{!}{\input{figures/ic_rule_diagram.tex}}
    \caption{A closed attainable set $\mathcal X=[u_1,v_1]\cup [u_2,v_2]$ and its induced IC mechanism. Demands in $\mathcal X$ receive their exact allocation, while the gap $(v_1,u_2)$ creates a jump at $\tau_1=(v_1+\alpha u_2)/(1+\alpha)$.}
    \label{fig:attainable_set_mechanism}
\end{figure}
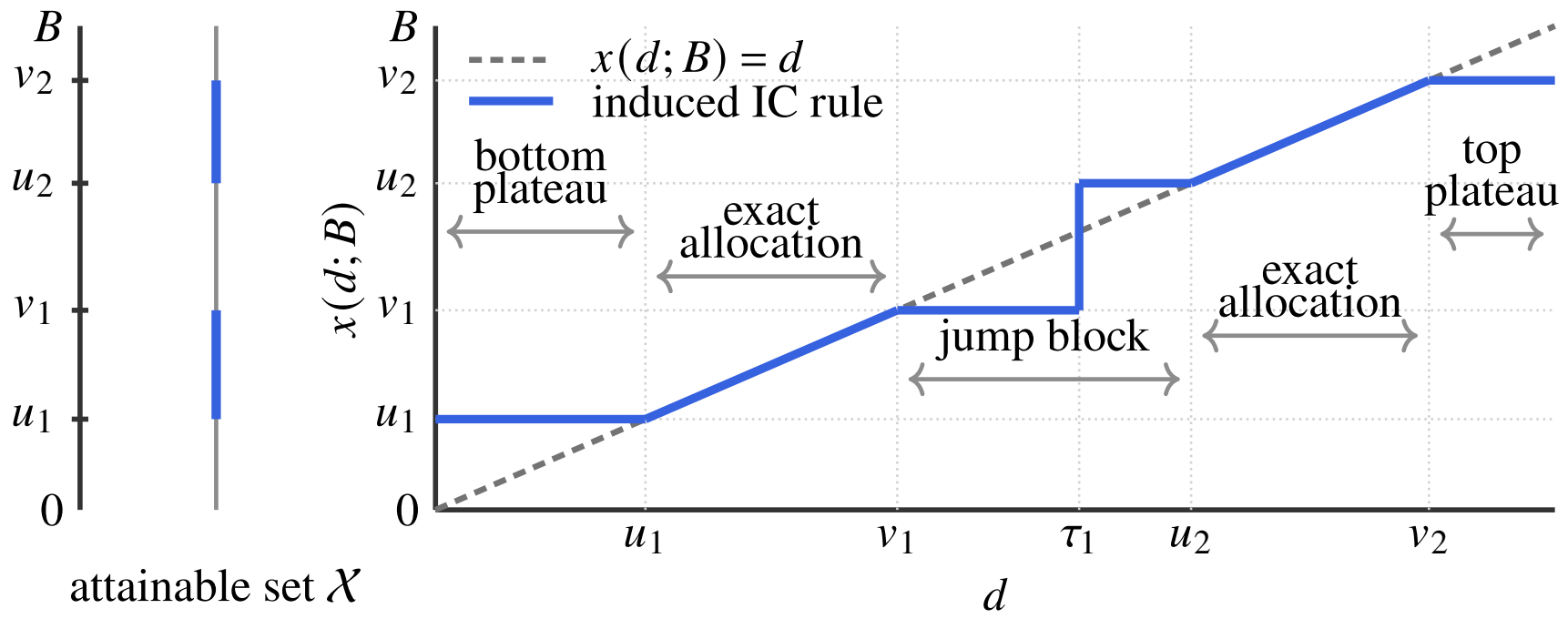

\noindent By \cref{thm:IC-structure}, any nonempty closed attainable set $\mathcal X\subseteq[0,B\wedge1]$ induces an IC rule by setting $x(d;B) = \argmax_{x \in \X} u(x,d;\alpha)$.  The resulting rule is weakly increasing, assigns exact allocation at attainable demands, and jumps between neighboring endpoints across each gap, with arbitrary tie-breaking at the cutoffs. Figure \ref{fig:attainable_set_mechanism} illustrates this construction for a given attainable set.

\begin{proof}
Fix an allocation rule $x(\cdot;B)$ and let $\mathcal X := \cl(x([0,1];B))$ denote its closed attainable set. By continuity of $u(\cdot,d;\alpha)$, incentive compatibility is equivalent to $x(d;B) \in \argmax_{y\in\mathcal X} u(y,d;\alpha)$ for every $d \in [0,1]$.

First suppose that $x(\cdot;B)$ is incentive compatible and let $m := \inf \X$ and $C := \sup X$. Since $\X$ is nonempty and closed, $m, C \in \X$. If $d < m$ then every attainable allocation satisifies $y > d$ and $u(y,d;\alpha) = (1 + \alpha)d - \alpha y$, which is decreasing in $y$. Hence, $x(d;B) = m$.  Similarly, if $d > C$ then every attainable allocation satisfies $y < d$ and $u(y,d;\alpha) = y$ which is increasing in $y$, so $x(d;B) = C$.  Moreover, if $d \in \X$ then the single-peaked utility $u(\cdot,d;\alpha)$ is uniquely maximized at $y = d$, implying that $x(d;B) = d$.

It remains to consider a gap $(v_j, u_{j+1})$ of $\X$. For $d \in (v_j, u_{j+1})$ the utility is increasing over attainable allocations below $d$, and decreasing over those above $d$, so only $v_j$ and $u_{j+1}$ can be optimal. Their utilities are $u(v_j,d;\alpha)=v_j$ and $u(u_{j+1},d;\alpha)=(1+\alpha)d-\alpha u_{j+1}.$ These are equal precisely when $d = \tau_j = \frac{v_j + \alpha u_{j+1}}{1 + \alpha}$. Hence, $v_j$ is optimal for $d < \tau_j$, $u_{j+1}$ is optimal for $d > \tau_j$, and either is optimal at $d = \tau_j$. This gives the stated rule.

Conversely, suppose $x(\cdot;B)$ is constructed from a nonempty closed set $\mathcal X$ as in the theorem. The preceding comparisons show that for every $d\in[0,1]$, $x(d;B)$ maximizes $u(\cdot,d;\alpha)$ over $\mathcal X$.
Since every deviation induces an allocation in $\mathcal X$, truthful reporting is optimal. Hence $x(\cdot;B)$ is incentive compatible.
\end{proof}

%% file: figures/ic_rule_diagram.tex
\begin{tikzpicture}[
    x=1.25cm,
    y=1.00cm,
    every node/.style={
        paper figure text,
        font=\normalsize
    },
    axis/.style={
        paper axis,
        line width=1.2pt
    },
    gridline/.style={
        paper grid,
        line width=0.55pt
    },
    diag/.style={
        paper benchmark,
        line width=1.4pt
    },
    rule/.style={
        paper primary,
        line width=2.0pt
    },
    interval/.style={
        paper primary,
        line width=2.2pt
    },
    annotation/.style={
        <->,
        paper auxiliary,
        line width=1.0pt
    },
    arrowlabel/.style={
        paper tick label,
        font=\normalsize
    },
]

\def\B{4.0}

\def\uone{0.75}
\def\vone{1.65}
\def\utwo{2.70}
\def\vtwo{3.55}
\def\done{2.30}

\def\leftx{0}
\def\rightx{2.35}

\def\xstretch{1.85}

\def\gap{0.04}

\begin{scope}[shift={(\leftx,0)}]

    \draw[axis] (0,0) -- (0,\B);

    \node[left] at (0,0) {$0$};
    \node[left] at (0,\B) {$B$};

    \foreach \y/\lab in {
        \uone/$u_1$,
        \vone/$v_1$,
        \utwo/$u_2$,
        \vtwo/$v_2$
    }{
        \draw[axis, line width=1.2pt]
            (-0.055,\y) -- (0.055,\y);
        \node[left] at (-0.055,\y) {\lab};
    }

    \draw[
        paper auxiliary,
        line width=1.0pt
    ]
        (0.9,0) -- (0.9,\B);

    \draw[interval]
        (0.9,\uone) -- (0.9,\vone);

    \draw[interval]
        (0.9,\utwo) -- (0.9,\vtwo);

    \node[below]
        at (0.9,-0.35)
        {attainable set $\mathcal X$};

\end{scope}

\begin{scope}[
    shift={(\rightx,0)},
    xscale=\xstretch
]

    \draw[axis] (0,0) -- (0,\B);
    \draw[axis] (0,0) -- (\B,0);

    \node[left] at (0,0) {$0$};
    \node[left] at (0,\B) {$B$};

    \node[below]
        at (\B/2,-0.43)
        {$d$};

    \node[rotate=90]
        at (-0.33,\B/2)
        {$x(d;B)$};

    \foreach \x in {
        \uone,
        \vone,
        \done,
        \utwo,
        \vtwo
    }{
        \draw[gridline]
            (\x,0) -- (\x,\B);
    }

    \foreach \y in {
        \uone,
        \vone,
        \utwo,
        \vtwo
    }{
        \draw[gridline]
            (0,\y) -- (\B,\y);
    }

    \foreach \x/\lab in {
        \uone/$u_1$,
        \vone/$v_1$,
        \done/$\tau_1$,
        \utwo/$u_2$,
        \vtwo/$v_2$
    }{
        \node[below]
            at (\x,0)
            {\lab};
    }

    \foreach \y/\lab in {
        \uone/$u_1$,
        \vone/$v_1$,
        \utwo/$u_2$,
        \vtwo/$v_2$
    }{
        \node[left]
            at (0,\y)
            {\lab};
    }

    \draw[diag]
        (0,0) -- (\B,\B);


    \draw[rule]
        (0,\uone) -- (\uone,\uone);

    \draw[rule]
        (\uone,\uone) -- (\vone,\vone);

    \draw[rule]
        (\vone,\vone) -- (\done,\vone);

    \draw[rule]
        (\done,\vone) -- (\done,\utwo);

    \draw[rule]
        (\done,\utwo) -- (\utwo,\utwo);

    \draw[rule]
        (\utwo,\utwo) -- (\vtwo,\vtwo);

    \draw[rule]
        (\vtwo,\vtwo) -- (\B,\vtwo);

    \draw[diag]
        (0.12,3.72) -- (0.42,3.72);

    \node[anchor=west]
        at (0.50,3.72)
        {$x(d;B)=d$};

    \draw[rule]
        (0.12,3.38) -- (0.42,3.38);

    \node[anchor=west]
        at (0.50,3.38)
        {induced IC rule};


    \draw[annotation]
        ({0.00+\gap},2.30)
        --
        ({\uone-\gap},2.30);

    \node[
        arrowlabel,
        above=0.5pt,
        align=center
    ]
        at ({0.5*\uone},2.30)
        {\shortstack{bottom\\[-3pt]plateau}};

    \draw[annotation]
        ({\uone+\gap},1.93)
        --
        ({\vone-\gap},1.93);

    \node[
        arrowlabel,
        above=0.5pt
    ]
        at ({(\uone+\vone)/2},1.93)
        {{\shortstack{exact\\[-3pt]allocation}}};

    \draw[annotation]
        ({\vone+\gap},1.08)
        --
        ({\utwo-\gap},1.08);

    \node[
        arrowlabel,
        above=0.5pt
    ]
        at ({(\vone+\utwo)/2},1.08)
        {jump block};

    \draw[annotation]
        ({\utwo+\gap},1.44)
        --
        ({\vtwo-\gap},1.44);

    \node[
        arrowlabel,
        above=0.5pt
    ]
        at ({(\utwo+\vtwo)/2},1.44)
        {{\shortstack{exact\\[-3pt]allocation}}};

    \draw[annotation]
        ({\vtwo+\gap},2.28)
        --
        ({\B-\gap},2.28);

    \node[
        arrowlabel,
        above=0.5pt,
        align=center
    ]
        at ({(\vtwo+\B)/2},2.28)
        {\shortstack{top\\[-3pt]plateau}};

\end{scope}

\end{tikzpicture}

%% file: short_parts/cap_optimality.tex
\section{Optimality of Cap Mechanisms}
\label{sec:optimal-ic}

The previous section characterized the structure of incentive compatible mechanisms. While this gives a complete description of the feasible class of IC mechanisms, it also highlights the difficulty of the planner's optimization problem. In general, the planner must optimize over the closed attainable set $\mathcal X$ that determines the IC rule. Equivalently, the planner chooses which gaps to create in the allocation space, and each such gap induces a jump block.
Here we identify sufficient conditions under which the optimal allocation mechanism is simple and has no jumps. We focus on {\em cap mechanisms}, under which the planner chooses a scalar $C \in [0,B\wedge 1]$ and allocates
$
\label{eq:cap_mechanism}
x(d;B) = \min\{d,C\}.
$
Cap mechanisms are incentive compatible via \cref{thm:IC-structure}, easy to implement, and interpretable. We show that in two-stage problems, cap mechanisms are optimal whenever the overage penalty $\alpha_i$ is sufficiently small.

We first note that cap mechanisms are always optimal at the final stage. Since the fill rate is weakly increasing in the allocation, a pointwise optimal feasible allocation (see \cref{eq:optimal_recursion}) is $
x_n^*(d_n;B)=\min\{d_n,B \wedge 1\}$.
This rule is a cap mechanism with cap $C=B \wedge 1$, and it is incentive compatible by \cref{thm:IC-structure}. We therefore ask whether cap optimality persists at earlier stages, where the planner must trade off the current fill rate against the future value.

\begin{theorem}
\label{thm:two_stage_cap}
Consider a two-stage problem with initial budget \(B_1>0\) and define
\[
\bar\alpha_1(B_1) :=
\inf_{0\le t<c\le B_1\wedge1 \mid f_1(t)>0}
\frac{
    \frac{1}{c-t}\int_t^c f_1(u)\,du
}{
    f_1(t)
}.
\]
If \(0<\alpha_1\le \bar\alpha_1(B_1)\), then there exists an optimal deterministic incentive-compatible first-stage mechanism of cap form: $x_1^*(d;B_1)=\min\{d,C\}$ for some \(C\in[0,B_1\wedge1]\). Moreover, if \(f_1\) is continuous and weakly increasing on \([0,B_1\wedge1]\), then \(\bar\alpha_1(B_1)=1\).
\end{theorem}

Before proving \cref{thm:two_stage_cap} we start with the following lemma.

\begin{lemma}
\label{lem:kappa_cap}
For any $i \in [n-1]$ and remaining budget $B_i$, suppose that the continuation value $V_{i+1}:[0,B_i]\to[0,1]$ is absolutely continuous and nondecreasing. Suppose there exists $\kappa \geq 1$ such that
\begin{equation}
\label{eq:kappa_secant_bound}
V_{i+1}(B_i-a)-V_{i+1}(B_i-z)
\leq
\kappa V_{i+1}'(B_i-z)(z-a)
\end{equation}
for every $a\in[0,b_i)$ and almost every $z\in(a,b_i)$. Define
\[
\bar\alpha^\kappa_{\mathrm{avg}}(B_i;F_i)
:=
\inf_{0\le t<c\le B_i\wedge1 \mid f_i(t)>0}
\frac{1}{\kappa}
\frac{
    \frac{1}{c-t}\int_t^c f_i(u)\,du
}{
    f_i(t)
}.
\]
If \(0<\alpha_i\le\bar\alpha^\kappa_{\mathrm{avg}}(B_i;F_i)\), then
there exists an optimal deterministic IC stage-\(i\) rule of cap form
\(
x_i^*(d_i;B_i)=\min\{d_i,C^*\}
\)
for some \(C^*\in[0,B_i\wedge1]\).
\end{lemma}

The proof of \cref{lem:kappa_cap} relies on the following lemma, showing that any IC rule can be approximated by IC rules whose attainable sets have finitely many gaps.

\begin{lemma}
\label{lem:finite-gap-approx}
For any $\alpha > 0$, $B > 0$, and nonempty closet set $\X \subseteq [0, B \wedge 1]$ let $x$ be the deterministic IC rule induced by $\X$ via \cref{thm:IC-structure} with ties broken by choosing the smaller minimizer. Further let $m = \inf \X$, $C = \sup \X$, and write $(m,C)\setminus\mathcal X=\bigcup_{j\ge1}(v_j,u_{j+1})$ as the countable disjoint union of gaps of \(\mathcal X\). 
For each integer \(K\ge1\), define \(\mathcal X^K:= \mathcal X\cup \bigcup_{j:\,u_{j+1}-v_j\le1/K}[v_j,u_{j+1}]\), and let \(x^K\) be the IC rule induced by \(\mathcal X^K\), again with
ties broken by choosing the smaller maximizer. Then:
\begin{enumerate}
    \item \(\mathcal X^K\) is closed, and
    \((m,C)\setminus\mathcal X^K\) has only finitely many gaps;

    \item \(x^K(d)\to x(d)\) for \(F\)-almost every \(d\in[0,1]\);

    \item for any continuous continuation function
    \(W:[0,B]\to[0,1]\), if
    \(J(y):=\E_{d\sim F}
    [\min\{\beta(y(d),d),W(B-y(d))\}]\),
    then \(J(x^K)\to J(x)\).
\end{enumerate}
\end{lemma}

\begin{proof}
The gaps remaining in \((m,C)\setminus\mathcal X^K\) are the
original gaps \((v_j,u_{j+1})\) with length
\(u_{j+1}-v_j>1/K\). Since they are disjoint and contained in
\([0,B\wedge1]\), there are finitely many. Hence
\(\mathcal X^K=[m,C]\setminus
\bigcup_{\ell=1}^N(r_\ell,s_\ell)\) is closed.
Moreover, every point in \(\mathcal X^K\) is within distance
\(1/K\) of \(\mathcal X\), since every added point lies in a
gap of length at most \(1/K\) whose endpoints belong to
\(\mathcal X\).

Fix \(d\in[0,1]\) such that the maximizer of
\(u(\cdot,d;\alpha)\) over \(\mathcal X\) is unique, and write
\(y_K:=x^K(d)\). Every subsequence of \((y_K)\) has a further
convergent subsequence; take one with \(y_{K_\ell}\to\bar y\).
Since every point of \(\mathcal X^{K_\ell}\) is within distance
\(1/K_\ell\) of \(\mathcal X\), the closedness of \(\mathcal X\)
gives \(\bar y\in\mathcal X\). For any \(z\in\mathcal X\),
optimality over \(\mathcal X^{K_\ell}\) gives \(u(y_{K_\ell},d;\alpha)\ge u(z,d;\alpha)\).
Taking limits, by continuity of \(u(\cdot,d;\alpha)\), gives
\(u(\bar y,d;\alpha)\ge u(z,d;\alpha)\), so uniqueness implies
\(\bar y=x(d)\).
By \cref{thm:IC-structure}, nonuniqueness occurs only at the
countably many gap cutoffs. Since \(F\) is atomless, these
have \(F\)-measure zero. Hence \(x^K(d)\to x(d)\) for
\(F\)-almost every \(d\).

For each \(d\), the term
\(\min\{\beta(y(d),d),W(B-y(d))\}\) is bounded by \(1\) and
continuous in \(y\) at every \(d>0\). Since \(F\) is atomless,
the value \(d=0\) has probability zero. Hence dominated
convergence implies \(J(x^K)\to J(x)\).
\end{proof}

\noindent Using this we now show \cref{lem:kappa_cap}.

\begin{proof}[Proof of \cref{lem:kappa_cap}]
Fix the nonterminal stage \(i\) and remaining budget \(B_i\),
and an arbitrary deterministic stage-\(i\) IC rule \(x_i\).
Let
\(\mathcal X:=\cl(x_i([0,1];B_i))
\subseteq[0,B_i\wedge1]\)
be its closed attainable set, and define
\[
J_i(x_i)
:=
\E_{d_i\sim F_i}
\left[
\min\{
    \beta(x_i(d_i;B_i),d_i),
    V_{i+1}(B_i-x_i(d_i;B_i))
\}
\right].
\]
Because \(F_i\) has a continuous density, cutoff types with
multiple utility-maximizing attainable allocations have probability zero. We may therefore use downward tie-breaking without
changing any objective value.

\paragraph{Step 1: The infimum of the attainable set can be taken to be zero.}
Let \(m:=\inf\mathcal X\). If \(m=0\), there is nothing to prove
in this step. Suppose \(m>0\), define
\(\mathcal X':=[0,m]\cup\mathcal X\), and let \(x'\) be the
IC rule induced by maximizing \(u(\cdot,d;\alpha_i)\) over
\(\mathcal X'\), with ties broken by choosing the smaller
maximizer.

For \(d\le m\), the rule \(x'\) assigns \(x'(d;B_i)=d\),
while the original rule assigns \(x_i(d;B_i)=m\) for \(d<m\)
and \(x_i(m;B_i)=m\). Hence the payoff under \(x'\) is
\(V_{i+1}(B_i-d)\ge V_{i+1}(B_i-m)\) by monotonicity of
\(V_{i+1}\), and is weakly larger than the payoff from
assigning \(m\) to such demand values.

For \(d>m\), adding the interval \([0,m]\) does not change
the chosen allocation. Among all points in \([0,m]\),
\(u(\cdot,d;\alpha_i)\) is uniquely maximized at \(m\),
and \(m\) was already in \(\mathcal X\). Therefore
\(x'(d;B_i)=x_i(d;B_i)\) for almost every \(d>m\).
Hence \(J_i(x')\ge J_i(x_i)\). Thus, without loss of
generality, we may restrict attention to IC rules whose
closed attainable set has infimum zero.

\paragraph{Step 2: It is enough to consider attainable sets with finitely many gaps.}
By \cref{lem:finite-gap-approx}, any IC rule whose attainable
set has infimum zero can be approximated in value by IC rules
whose attainable sets have finitely many gaps. We establish
cap dominance for these rules first and return to arbitrary
attainable sets in Step 5. Fix such a finite-gap rule \(x_i\),
with closed attainable set \(\mathcal X\) satisfying
\(\inf\mathcal X=0\).

\paragraph{Step 3: Filling the rightmost gap weakly improves the objective.}
Let \(C:=\sup\mathcal X\). If \(\mathcal X=[0,C]\), then
\(x_i(d;B_i)=\min\{d,C\}\), and the rule is already a cap.
Otherwise, let \((v_j,u_{j+1})\) be the rightmost gap of
\(\mathcal X\), meaning that \(u_{j+1}\) is maximal among
all right endpoints of gaps. Then
\([u_{j+1},C]\subseteq\mathcal X\).

For each \(z\in[v_j,u_{j+1}]\), define a new closed attainable
set \(\mathcal Y(z):=(\mathcal X\cap[0,v_j])\cup[z,C]\),
and and let \(x_z\) be the IC rule induced by maximizing
\(u(\cdot,d;\alpha_i)\) over \(\mathcal Y(z)\). When \(z=u_{j+1}\), this rule coincides
with the original rule on and to the right of the gap.
When \(z=v_j\), the gap \((v_j,u_{j+1})\) has been filled.
Define \(t(z):=\frac{(v_j+\alpha_i z)}{(1+\alpha_i)}\).
Note that for demands \(d\le v_j\), the induced IC rule is
unchanged from the original rule, that is,
\(x_z(d;B_i)=x_i(d;B_i)\).
For \(d>v_j\), the induced rule on \((v_j,1]\) is
\[
x_z(d;B_i)=
\begin{cases}
v_j, & d\in(v_j,t(z)],\\
z, & d\in(t(z),z],\\
d, & d\in(z,C],\\
C, & d\in(C,1].
\end{cases}
\]

Using the expression for \(x_z\), the terms for \(d\le v_j\)
and \(d>C\) are independent of \(z\). Therefore, to compare
the rules \(x_z\), it is enough to study the \(z\)-dependent
part
\[
\begin{aligned}
\mathcal H(z)
={}&
\int_{(v_j,t(z)]}
\min\left\{
    \frac{v_j}{d},
    V_{i+1}(B_i-v_j)
\right\}\,dF_i(d)
+
V_{i+1}(B_i-z)\bigl(F_i(z)-F_i(t(z))\bigr)
+
\int_{(z,C]}V_{i+1}(B_i-d)\,dF_i(d).
\end{aligned}
\]
The three terms correspond to the demands assigned the
lower endpoint, the demands assigned the upper endpoint,
and the demands receiving exact allocation. Moving the upper
endpoint to the left fills the gap, so we compare these three
contributions along that movement.

We show that \(\mathcal H\) is weakly decreasing on
\([v_j,u_{j+1}]\). Since \(V_{i+1}\) is absolutely continuous
and \(F_i\) has a continuous density, \(\mathcal H\) is
absolutely continuous. At almost every
\(z\in(v_j,u_{j+1})\) at which \(V_{i+1}\) is differentiable
at \(B_i-z\), writing \(t=t(z)\) and
\(t'(z)=\alpha_i/(1+\alpha_i)\), differentiation gives
\[
\mathcal H'(z)
=
\frac{\alpha_i}{1+\alpha_i}f_i(t)
\left[
\min\left\{\frac{v_j}{t},V_{i+1}(B_i-v_j)\right\}
-
V_{i+1}(B_i-z)
\right]
-
V_{i+1}'(B_i-z)\bigl(F_i(z)-F_i(t)\bigr).
\]
The first term captures the payoff effect of shifting the
cutoff between the two endpoint allocations. The second
captures the loss of continuation value as the upper endpoint
allocation increases.

Let \(\lambda(z):=V_{i+1}'(B_i-z)\), which is nonnegative
for almost every \(z\) because \(V_{i+1}\) is nondecreasing.
Also,
\(\min\{v_j/t,V_{i+1}(B_i-v_j)\}
\le V_{i+1}(B_i-v_j)\).
By the \(\kappa\)-secant condition in
\cref{eq:kappa_secant_bound}, applied with \(a=v_j\),
and the identity \(z-v_j=(1+\alpha_i)(z-t)\),
\[
\mathcal H'(z)
\le
\lambda(z)
\left[
\kappa\alpha_i f_i(t)(z-t)-\int_t^z f_i(u)\,du
\right].
\]
If \(f_i(t)=0\), then the bracket is nonpositive.
If \(f_i(t)>0\), then, since
\(0\le t<z\le u_{j+1}\le B_i\wedge1\),
the pair \((t,z)\) is admissible in the definition of
\(\bar\alpha_{\mathrm{avg}}^{\,\kappa}(B_i;F_i)\), and
\(\alpha_i\le
\bar\alpha_{\mathrm{avg}}^{\,\kappa}(B_i;F_i)\)
implies that the bracket is nonpositive. Thus
\(\mathcal H'(z)\le0\) for almost every
\(z\in(v_j,u_{j+1})\). Since \(\mathcal H\) is absolutely
continuous, \(\mathcal H\) is weakly decreasing on
\([v_j,u_{j+1}]\).
It follows that \(J_i(x_{v_j})\ge J_i(x_{u_{j+1}})\),
so filling the rightmost gap weakly improves the objective.

\paragraph{Step 4: Iterating the gap-filling step produces a cap rule.}
Repeat Step 3 for each of the finitely many gaps. Each step
weakly improves the objective and removes the rightmost
remaining gap, yielding \([0,C^*]\) for some
\(C^*\in[0,B_i\wedge1]\). The induced IC rule is
the cap \(d\mapsto\min\{d,C^*\}\). Therefore every finite-gap
IC rule with infimum zero is weakly dominated by a cap rule.

\paragraph{Step 5: Passing to arbitrary closed attainable sets.}
Return to the arbitrary IC rule \(x_i\) from the beginning
of the proof, after the normalization in Step 1. By Step 2,
there exists a sequence of finite-gap IC rules \(x_i^K\)
such that \(J_i(x_i^K)\to J_i(x_i)\). By Step 4, for each
\(K\), there is a cap rule \(y^K(d)=\min\{d,C_K\}\), with
\(C_K\in[0,B_i\wedge1]\), such that
\(J_i(y^K)\ge J_i(x_i^K)\).

Define the cap objective
\[
\Psi(C)
:=
\E_{d_i\sim F_i}
\left[
\min\{
    \beta(\min\{d_i,C\},d_i),
    V_{i+1}(B_i-\min\{d_i,C\})
\}
\right],
\qquad C\in[0,B_i\wedge1].
\]
The function \(\Psi\) is continuous by dominated convergence.
Since \([0,B_i\wedge1]\) is compact, some subsequence
\(C_{K_\ell}\) converges to a limit
\(C^\star\in[0,B_i\wedge1]\). By continuity,
\(\Psi(C^\star)
=\lim_{\ell\to\infty}J_i(y^{K_\ell})
\ge\lim_{\ell\to\infty}J_i(x_i^{K_\ell})
=J_i(x_i)\).
Thus the original IC rule \(x_i\) is weakly dominated by
the cap rule \(d\mapsto\min\{d,C^\star\}\).
Finally, \(\Psi\) attains its maximum on \([0,B_i\wedge1]\).
Since every deterministic stage-\(i\) IC rule is weakly
dominated by some cap rule, a maximizing cap is optimal.
\end{proof}

Finally we can apply \cref{lem:kappa_cap} to prove
\cref{thm:two_stage_cap}.

\begin{proof}[Proof of \Cref{thm:two_stage_cap}]
At stage \(2\), the optimal cap rule described above has value
\(V_2(B_2)=
\E_{d_2\sim F_2}[\beta(\min\{d_2,B_2\},d_2)]\).
For every realized demand \(d_2\), the expression inside
the expectation is nondecreasing and concave as a function
of \(B_2\). Taking expectations preserves these properties,
so \(V_2\) is nondecreasing and concave on \([0,B_1]\).
Moreover, the expression inside the expectation is bounded
by \(1\) and continuous in \(B_2\), so \(V_2\) is continuous.
Since every continuous concave function on a closed interval
is absolutely continuous
\citep[Section~4.9.3]{MakarovPodkorytov2013},
\(V_2\) is absolutely continuous on the interval \([0,B_1]\).

We can therefore apply \Cref{lem:kappa_cap} to the first-stage
problem with \(i=1\), continuation value \(V_{i+1}=V_2\),
and \(\kappa=1\). Indeed, concavity gives
\(V_2(B_1-a)-V_2(B_1-z)
\le V_2'(B_1-z)(z-a)\)
for every \(a\in[0,B_1\wedge1)\) and almost every
\(z\in(a,B_1\wedge1)\), and
\(\bar\alpha_{\mathrm{avg}}^{\,1}(B_1;F_1)
=\bar\alpha_1(B_1)\).
Hence, when \(0<\alpha_1\le\bar\alpha_1(B_1)\),
\Cref{lem:kappa_cap} guarantees the existence of an optimal
deterministic IC first-stage mechanism of cap form.

Finally, suppose that \(f_1\) is continuous and weakly
increasing on \([0,B_1\wedge1]\). For every admissible
\(t<c\),
\(\frac{1}{(c-t)}\int_t^c f_1(u)\,du\ge f_1(t)\),
so \(\bar\alpha_1(B_1)\ge1\). Conversely, fix any
\(t\in(0,B_1\wedge1)\). By the standing positive-density
assumption, \(f_1(t)>0\), and continuity implies that
the ratio in the definition of \(\bar\alpha_1(B_1)\)
converges to \(1\) as \(c\downarrow t\). Thus
\(\bar\alpha_1(B_1)=1\).
\end{proof}

The proof of \cref{thm:two_stage_cap} also explains why small overage penalties make cap mechanisms optimal. By \cref{thm:IC-structure}, any over-allocation region in a non-cap mechanism must be paired with an under-allocation region, creating a gap in the attainable set.
When \(\alpha\) is small, the over-allocation region is relatively large; if the next-stage gain from saving additional inventory is sufficiently bounded, filling each gap weakly improves the objective, and hence a cap mechanism weakly dominates any non-cap mechanism. Cap mechanisms need not be optimal in general, however, and the two-stage argument does not extend directly to \(n\)-stage problems because the incentive-compatible Bellman operator need not preserve the concavity required for backward induction.

Finally, we note that the cap-optimality condition in \Cref{thm:two_stage_cap} only gives a sufficient condition. However, cap mechanisms need not be optimal when the overage penalties $\alpha_i$ are large. Consider a two-stage problem ($n = 2$) with initial budget $B_1 = 0.45$. The overage penalty for the first stage is \(\alpha_1=10\). The value of \(\alpha_2\) is irrelevant for this example because the final-stage optimal rule is \(x_2(d_2;B_2)=\min\{d_2,B_2\}\), independently of \(\alpha_2\).  The second stage demand distribution is $F_2 = \text{Beta}(5,0.9)$, and the first stage distribution is $F_1 = \frac{1}{2}\text{Beta}(6,34) + \frac{1}{2}\text{Beta}(15,15)$.  We note that the second-stage distribution places substantial mass on large demands, whereas the first-stage distribution places mass at both low and intermediate demand values.

Numerically, the best cap mechanism has cap $C^*\approx0.169$ and achieves an objective value of approximately $0.35$. However, the attainable set $\mathcal X^{\mathrm{jump}}=[0,0.071]\cup\{0.176\}$ induces a deterministic IC mechanism by \cref{thm:IC-structure}, with cutoff
$\tau=(0.071+\alpha_1\cdot0.176)/(1+\alpha_1)\approx0.167$.
Evaluating this feasible one-jump IC mechanism numerically gives an objective value of approximately \(0.37\), which is strictly larger than the value achieved by the best cap mechanism. Thus, in this instance, the best cap mechanism is strictly dominated by a feasible one-jump IC mechanism.

%% file: short_parts/approximation.tex
\section{Approximation Methods for the Optimal Incentive-Compatible Mechanism}
\label{sec:approximation}

The previous sections characterize deterministic IC mechanisms through their closed attainable sets $\X$ and identify conditions under which the optimal mechanism is a cap mechanism. Outside of these regimes, however, to find the optimal fair allocation the planner must optimize over an infinite-dimensional class of attainable sets $\X$.  Motivated by this, we develop an approximation framework for computing the optimal incentive compatible mechanism by discretizing the set of feasible allocation levels. We then show that the resulting discrete problem can be solved exactly by reducing it to a longest path problem on a carefully constructed directed acyclic graph.

We start by discretizing the planner's allocation space. Fix a grid size $\epsilon>0$ and define the budget grid $G_\epsilon=\{0,\epsilon,2\epsilon,\ldots,B_1\}$, where we assume for convenience that $B_1$ is a multiple of $\epsilon$.  For any $B \in G_\epsilon$ define the feasible allocation grid $A_\epsilon(B):=\{a\in G_\epsilon:0\leq a\leq B\wedge1\}$. This ensures that whenever $a \in A_\epsilon(B)$ then $B - a \in G_\epsilon$, so the discretized problem is closed over the finite budget grid.
Because agents know the mechanism, restricting allocations to $A_\epsilon(B)$ also changes the incentive-compatibility constraints. Nevertheless, a similar structure to \cref{thm:IC-structure} continues to hold: an IC rule assigns each demand its preferred attainable allocation, but the attainable set is now finite. This yields the following characterization of discrete IC mechanisms.

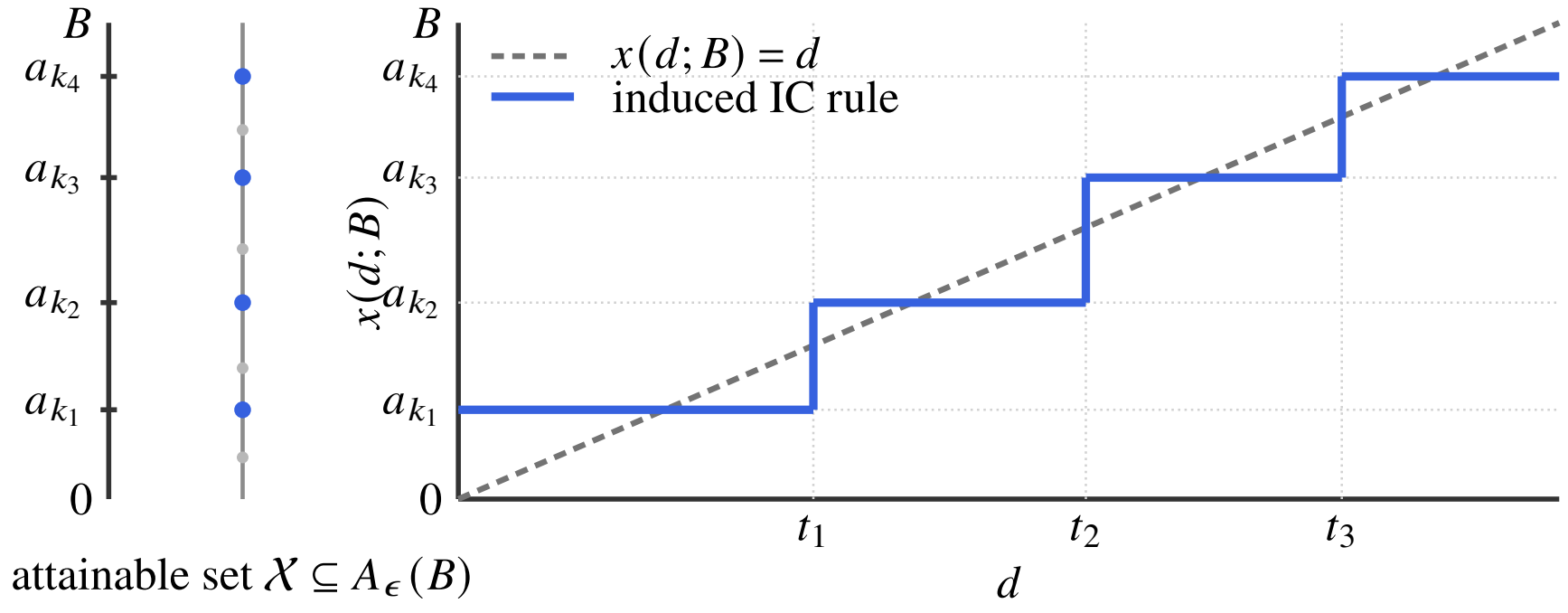
\begin{figure}[!t]
    \centering
    \resizebox{0.7\textwidth}{!}{\input{figures/ic_rule_discrete}}
\caption{Discrete analogue of the IC characterization for $\alpha=1.5$. The left panel shows a realized allocation set $\mathcal X=\{a_{k_1}<a_{k_2}<a_{k_3}<a_{k_4}\}\subseteq A_\epsilon(B)$. The right panel shows the induced deterministic IC rule, which is a step function with cutoffs $t_j=\frac{a_{k_j}+1.5\,a_{k_{j+1}}}{2.5}$.}
    \label{fig:attainable_set_mechanism_discrete}
\end{figure}

\begin{proposition}
\label{prop:finite_ic_characterization}
Fix $\alpha > 0$, a budget $B \in G_\epsilon$, and the finite allocation set $A_\epsilon(B)$. A deterministic allocation rule $x(\cdot;B) : [0,1] \rightarrow A_\epsilon(B)$ is incentive compatible if and only if there exists a finite ordered set
\[
\mathcal X=\{a_{k_1}<\cdots<a_{k_K}\}\subseteq A_\epsilon(B)
\]
such that, defining $t_j:=\frac{a_{k_j}+\alpha a_{k_{j+1}}}{1+\alpha}$, for $j=1,\ldots,K-1,$ with $t_0=0$ and $t_K=1$, we have
\[
x(d;B)=a_{k_j},
\qquad
d\in(t_{j-1},t_j],\quad j=1,\ldots,K,
\]
where the first interval includes $d=0$ and tie-breaking at the cutoffs $t_j$ is arbitrary.
\end{proposition}
\noindent \noindent We omit the proof, which follows similarly to \cref{thm:IC-structure}. Figure~\ref{fig:attainable_set_mechanism_discrete} illustrates the construction of an IC allocation rule from a finite ordered attainable set.

At first glance, this characterization still leaves a difficult combinatorial optimization problem. For a grid with $k$ feasible allocation levels, directly optimizing over the realized allocation set would require searching over all $O(2^k)$ many subsets. We avoid this enumeration by exploiting the local structure of the IC cutoffs where the boundary between two adjacent realized allocations depends only on those two allocations. This locality allows us to formulate the finite mechanism design problem as a longest-path problem on a directed acyclic graph.

We now define the (modified) dynamic program after discretizing the allocation space. Let
$\IC_\epsilon(B;\alpha)$ denote the set of incentive-compatible rules whose allocations lie in $A_\epsilon(B)$:
\[
\IC_\epsilon(B;\alpha)
:=
\{x(\cdot;B):[0,1]\to A_\epsilon(B)\mid x(\cdot;B)\text{ is incentive compatible given }\alpha\}.
\]
Let $V_i^\epsilon(B)$ denote the optimal value at stage $i$ when the current budget is $B$ and all future allocations are restricted to the grid. We again use the fact that the discretized problem has a recursive definition:
\begin{align*}
V_n^\epsilon(B) 
& :=
\sup_{x\in \IC_\epsilon(B;\alpha_n)}
\E_{d_n\sim F_n}\left[\beta(x(d_n;B),d_n)\right], \\
V_i^\epsilon(B)
& :=
\sup_{x\in \IC_\epsilon(B;\alpha_i)}
\E_{d_i\sim F_i}
\left[
\min\left\{
\beta(x(d_i;B),d_i),
V_{i+1}^\epsilon(B-x(d_i;B))
\right\}
\right], \quad i = n-1, \ldots, 1.
\end{align*}
Thus, the recursion has the same form as the continuous dynamic program, except that the planner optimizes only over IC rules whose allocations belong to the finite set $A_\epsilon(B)$.

We now show how to compute $V_i^\epsilon(B)$ for a fixed stage $i$ and current budget $B \in G_\epsilon$.  Fix a stage $i$ and assume that $V_{i+1}^\epsilon$ has already been computed. For each $a\in A_\epsilon(B)$, define
\begin{equation}
\label{eq:gamma_def}
\Gamma_{i,B}(a,s,t)
:=
\int_{(s,t]}
\min\{\beta(a,d),V_{i+1}^\epsilon(B-a)\}\,dF_i(d),
\end{equation}
with the convention that $V_{n+1}^\epsilon(B-a)=1$. Moreover, for two realized allocation levels $a < a'$, we define the induced cutoff
\[
\tau_i(a,a') = \frac{a + \alpha_i a'}{1 + \alpha_i}.
\]

Now consider any discretized IC mechanism whose realized allocation levels are $a_{k_1} < \ldots < a_{k_K} \in A_\epsilon(B)$.  If $K = 1$ then the objective value of the mechanism is simply $\Gamma_{i,B}(a_{k_1},0,1)$. However, if $K \geq 2$, then the objective value of the mechanism is:
\begin{equation}
\label{eq:gamma_mech_weight}
\Gamma_{i,B}(a_{k_1}, 0, \tau_i(a_{k_1}, a_{k_2})) + \sum_{j=2}^{K-1} \Gamma_{i,B}(a_{k_j}, \tau_i(a_{k_{j-1}}, a_{k_j}), \tau_i(a_{k_j}, a_{k_{j+1}}))
+ \Gamma_{i,B}(a_{k_K}, \tau_i(a_{k_{K-1}}, a_{k_K}), 1).
\end{equation}
This corresponds to allocating $a_{k_1}$ for demand values in $[0, \tau_i(a_{k_1}, a_{k_2})]$, allocating $a_{k_2}$ for demand values in $(\tau_i(a_{k_1}, a_{k_2}), \tau_i(a_{k_2}, a_{k_3})]$, etc.
Therefore, we see that computing $V_i^\epsilon(B)$ is equivalent to choosing an increasing sequence of realized allocation levels that maximizes this expression.

We formulate this selection problem as a longest-path problem on a directed acyclic graph. Algorithm~\ref{alg:dag_stage_solver} constructs the graph for a fixed stage $i$ and budget $B$. Singleton nodes represent constant mechanisms, while pair nodes record two consecutive attainable allocation levels. The optimal value is obtained by finding a longest source-to-sink path in this graph.

\begin{figure}[t]
    \centering
    \input{figures/dag_diagram}
\caption{Directed Acyclic
Graph (DAG) representation of the discretized optimal IC mechanism problem for three allocation levels $a < a' < a''$, with singleton paths omitted.}
\label{fig:dag_ic_three_levels}
\end{figure}
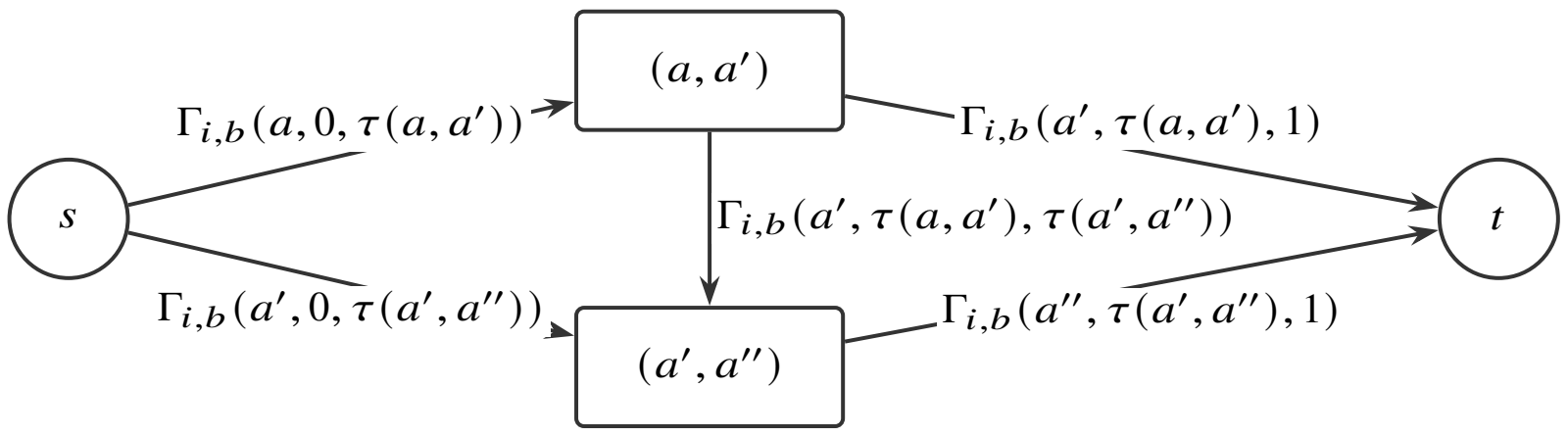

\begin{algorithm}[t]
\caption{Graph construction at stage $i$ and budget $B$}
\label{alg:dag_stage_solver}
\begin{algorithmic}[1]
\Require Allocation grid $A_\epsilon(B)$, interval payoffs $\Gamma_{i,B}$, and cutoffs $\tau_i$.

\State Define $\mathcal N=\{s,t\}\cup A\cup\{(a,a'):a,a'\in A,\ a<a'\}$ and initialize $\mathcal E \gets \varnothing$.

\ForAll{$a\in A_\epsilon(B)$}
    \State Add edge $s\to a$ with weight $0$, and edge $a\to t$ with weight $\Gamma_{i,B}(a,0,1)$.
\EndFor

\ForAll{$a,a'\in A_\epsilon(B)$ with $a<a'$}
    \State Add edge $s\to(a,a')$ with weight
    $\Gamma_{i,B}(a,0,\tau_i(a,a'))$, and an edge $(a,a')\to t$ with weight
    $\Gamma_{i,B}(a',\tau_i(a,a'),1)$.
    \ForAll{$a''\in A$ with $a''>a'$}
        \State Add edge $(a,a')\to(a',a'')$ with weight
        $\Gamma_{i,B}(a',\tau_i(a,a'),\tau_i(a',a''))$.
    \EndFor
\EndFor

\State \Return $\mathcal G=(\mathcal N,\mathcal E,w)$.
\end{algorithmic}
\end{algorithm}

See \cref{fig:dag_ic_three_levels} for an example of the graph constructed by Algorithm~\ref{alg:dag_stage_solver} with three allocation levels. Computing $V_i^\epsilon(B)$ reduces to finding a longest $s$--$t$ path in this graph, as established by the following proposition.

\begin{proposition}
\label{prop:dag_stage_solver}
Fix a stage $i$ and budget $B\in G_\epsilon$. The one-stage optimization problem defining $V_i^\epsilon(B)$ is equivalent to finding a longest $s$--$t$ path in the graph generated by Algorithm~\ref{alg:dag_stage_solver}.
\end{proposition}
\begin{proof}
A singleton path $s\to a\to t$ represents the constant IC mechanism with attainable set $\{a\}$ and value $\Gamma_{i,B}(a,0,1)$, equal to its path weight.

More generally, a path through pair nodes has the form \(s\to(a_{k_1},a_{k_2})\to\cdots\to(a_{k_{K-1}},a_{k_K})\to t\), where $a_{k_1}<\cdots<a_{k_K}$ and $K\geq2$, and hence represents the realized allocation set $\X=\{a_{k_1},\ldots,a_{k_K}\}$. By \cref{prop:finite_ic_characterization}, the induced IC mechanism has value in \cref{eq:gamma_mech_weight}, which equals the total path weight by construction. Conversely, every discretized IC mechanism determines exactly such a path. Thus maximizing over IC mechanisms is equivalent to finding the longest $s$--$t$ path.
\end{proof}

\cref{prop:dag_stage_solver} allows us to analyze the computation and storage complexity of solving for $V_1^\epsilon(B_1)$. For a fixed stage $i$ and budget $B\in G_\epsilon$, let $m(B):=|A_\epsilon(B)|$. The graph in \cref{prop:dag_stage_solver} has $O(m(B)^2)$ nodes and $O(m(B)^3)$ edges, since its transition edges are indexed by triples $a<a'<a''$. Because the graph is acyclic, its longest path can be computed in time linear in the number of nodes and edges \citep[Chapter~6]{kleinberg2006algorithm}. Although Algorithm~\ref{alg:dag_stage_solver} specifies the full graph, the longest-path computation can generate its edges as needed, storing only vertex labels and predecessor pointers. Thus, assuming that each $\Gamma_{i,B}(a,s,t)$ can be evaluated in constant time, a single stage-budget problem requires $O(m(B)^3)$ time and $O(m(B)^2)$ space. Aggregating over all $B\in G_\epsilon$ and $i\in[n]$ gives the following complexity bound.

\begin{proposition}
\label{prop:disc_complexity}
Let $M = B_1 / \epsilon$. Suppose that each interval payoff $\Gamma_{i,B}(a,s,t)$ can be evaluated in constant time. Then the discretized formulation above computes $V_1^\epsilon(B_1)$ in $O(nM^4)$ time, and $O(M^2)$ space.
\end{proposition}

Lastly we quantify the loss from restricting the current-stage allocation rules to the discrete set of allocations in $G_\epsilon$. We show that when the demand densities are bounded, $V_1^\epsilon(B_1)$ approximates the true optimal objective value $V_1(B_1)$ with an error of order $\sqrt{\epsilon}$.  Indeed, we show the following:

\begin{theorem}
\label{thm:grid_approx}
Let \(B_1<\infty\) and \(\epsilon\in(0,1]\) satisfy
\(B_1/\epsilon\in\mathbb N\). Suppose that, for each \(i\in[n]\),
\(F_i\) has full support on \([0,1]\) and admits a density \(f_i\)
satisfying \(f_i(d)\le \bar f<\infty\) for all \(d\in[0,1]\). Then
\[
0\le V_1(B_1)-V_1^\epsilon(B_1)
\le
\left[
n(\bar f+A)
+
\frac{r\sqrt A}{q-1}
\left(
\frac{q^n-1}{q-1}-n
\right)
\right]\sqrt{\epsilon},
\]
where
\(A=K_\alpha+\bar fB_1(6+3K_\alpha)\), \(K_\alpha=\max_{i\in[n]}\max\{\alpha_i,\alpha_i^{-1}\}\),
\(q=\sqrt{2+\bar fB_1}\), and
\(r=1+\bar f(1+B_1)\).
\end{theorem}
\noindent To show the result we start by proving two auxiliary lemmas. For any bounded continuation function $W$, define the unrestricted and grid-restricted Bellman operators by
\begin{align*}
    (\mathcal T_i W)(B)
& = \sup_{x\in\IC(B;\alpha_i)}
\E\left[ \min\left\{
\beta(x(d_i;B),d_i),
W(B-x(d_i;B))
\right\}
\right], \\
(\mathcal T_i^\epsilon W)(B)
& =
\sup_{x\in\IC_\epsilon(B;\alpha_i)}
\E\left[
\min\left\{
\beta(x(d_i;B),d_i),
W(B-x(d_i;B))
\right\}
\right] \text{ for } B \in G_\epsilon.
\end{align*}
Set \(V_{n+1}\equiv V_{n+1}^\epsilon\equiv1\), so that
\(V_i=\mathcal T_iV_{i+1}\) and
\(V_i^\epsilon=\mathcal T_i^\epsilon V_{i+1}^\epsilon\) for every
\(i\in[n]\). Throughout this proof and lemmas, we maintain the assumptions of
\cref{thm:grid_approx}.

We first show that $\T_i$ is H\"older continuous in the remaining budget.

\begin{lemma}
\label{lem:grid_lipschitz}
Let $D=1+\bar fB_1$, $L_{n+1}=0$, and $L_i=\bar f+D+\sqrt{1+D}\,L_{i+1}$ for $i=n,\ldots,1$. Then, for every $i\in[n+1]$ and $B,B'\in[0,B_1]$, \(|V_i(B)-V_i(B')|\le L_i\sqrt{|B-B'|}\).
\end{lemma}
\begin{proof}
Fix a stage $i$ and suppose $W:[0,B_1]\to[0,1]$ satisfies \(|W(B)-W(B')|\le L\sqrt{|B-B'|}\) for all $B,B'\in[0,B_1]$. We show that $\mathcal T_iW$ satisfies the same bound with \(L'=\bar f+D+\sqrt{1+D}\,L\), where \(D:=1+\bar fB_1\).
Fix $0 \leq B < B' \leq B_1$ and write $h:=B'-B$. 

We first assume $h \leq 1$.  For the lower comparison, take any $x\in\IC(B;\alpha_i)$. Since $x(d;B)\le B<B'$, the same allocation rule is feasible at budget $B'$. The fill-rate term is unchanged, and the continuation arguments differ by exactly $h$. Hence $(\mathcal T_i W)(B')\ge(\mathcal T_i W)(B)-L\sqrt h$.

For the upper comparison, take an $\eta$-optimal rule $x' \in \IC(B';\alpha_i)$ at budget $B'$ and let $\X = \cl(x'([0,1];B')) \subseteq [0,B'\wedge1]$ be its closed attainable set. Define the clipped attainable set by
$\X^- = \{\min(y,B) \mid y \in \X\}$
%
This is a nonempty closed subset of $[0,B\wedge1]$. Let $x^-$ be the IC rule induced by $\X^-$ via \cref{thm:IC-structure}. Then $x^- \in \IC(B;\alpha_i)$.  We first claim that $\E[\abs{x'(d_i;B') - x^-(d_i;B)}] \leq Dh$ for any $d_i$.

Indeed, if $\mathcal X\subseteq[0,B\wedge1]$, then $\mathcal X^-=\mathcal X$ and $x'=x^-$ for $F_i$-almost every $d_i$. If $\mathcal X\cap[0,B\wedge1]=\emptyset$, then every allocation under $x'$ lies in $[B,B']$, while $x^-\equiv B$, so $|x'(d_i;B')-x^-(d_i;B)|\le h$ for all $d_i$.
It remains to consider the case in which $\mathcal X$ has points both weakly below and strictly above $B$. Let \(v=\sup(\mathcal X\cap[0,B\wedge1])\) and \(u=\inf(\mathcal X\cap(B,B'])\).
Under $\mathcal X$, the relevant cutoff is $c=(v+\alpha_i u)/(1+\alpha_i)$. Under $\mathcal X^-$, the corresponding cutoff is $c^-=(v+\alpha_i B)/(1+\alpha_i)$. Hence
\(0\le c-c^-\le h\).
Outside $(c^-,c]$, clipping changes allocations by at most $h$. On $(c^-,c]$, the allocation difference is at most $B_1$. Since $f_i\le\bar f$, this interval has probability at most $\bar f h$. Therefore \(\E[|x'(d_i;B')-x^-(d_i;B)|]\le Dh\).

Write \(\Delta(d_i):=|x'(d_i;B')-x^-(d_i;B)|\). Since $\beta(\cdot, d_i) \in [0,1]$ and is $1/d_i$ Lipschitz in its first argument:
\[
\E[\abs{\beta(x'(d_i;B'),d_i) - \beta(x^-(d_i;B), d_i)}] \leq \Pr(d_i \leq \sqrt{h}) + \frac{\E[\Delta(d_i)]}{\sqrt{h}} \leq (\bar f + D) \sqrt{h}.
\]
Moreover, the next-stage budgets $B' - x'(d_i;B')$ and $B - x^-(d_i;B)$ differ by at most $h + \Delta(d_i)$.  Hence, by the assumed bound on $W$ and Jensen's inequality we have
\[
\E[\abs{W(B' - x'(d_i;B')) - W(B - x^-(d_i;B))}] \leq L\sqrt{1+D}\sqrt{h}.
\]
Using the $\eta$-optimality of $x'$ and feasibility of $x^-$ this gives \((\mathcal T_iW)(B')-(\mathcal T_iW)(B)\le(\bar f+D+L\sqrt{1+D})\sqrt h+\eta\).
%
%
Letting $\eta\downarrow0$ gives the upper comparison. Together with the lower comparison and the fact that $L'\ge L$, this proves \(|(\mathcal T_iW)(B')-(\mathcal T_iW)(B)|\le L'\sqrt h\) when $h\le1$. When $h>1$, the same inequality follows because both values lie in $[0,1]$ and $L'\ge1$.

Starting from $V_{n+1}\equiv1$ and $L_{n+1}=0$, backward induction proves the lemma.
\end{proof}

We next bound the approximation error introduced by restricting the allocations to the grid $G_\epsilon$.

\begin{lemma}
\label{lem:grid_operator_approx}
Fix a stage $i$ and suppose that $W:[0,B_1]\to[0,1]$ satisfies \(|W(B)-W(B')|\le L\sqrt{|B-B'|}\). Define $K_i=\max\{\alpha_i,\alpha_i^{-1}\}$ and $A_i=K_i+\bar fB_1(6+3K_i)$. Then, for every $\epsilon\in(0,1]$ and $B\in G_\epsilon$, 
$$0\le(\mathcal T_iW)(B)-(\mathcal T_i^\epsilon W)(B)\le(\bar f+A_i+L\sqrt{A_i})\sqrt{\epsilon}.$$
\end{lemma}

\begin{proof}
The result is immediate at $B=0$, so fix $B\in G_\epsilon$ with $B>0$. The lower bound is immediate, since every grid-feasible IC rule is feasible for the unrestricted problem, $(\mathcal T_i^\epsilon W)(B)\le(\mathcal T_iW)(B)$.
Fix $\eta>0$ and choose an $\eta$-optimal $x\in\IC(B;\alpha_i)$.
Let $\mathcal X=\cl(x([0,1];B))\subseteq[0,B\wedge1]$ be the closed attainable set of $x$. For each $y\in[0,B\wedge1]$, let $q_\epsilon(y)$ be a nearest point in $A_\epsilon(B)$, with ties broken downward. Then $|q_\epsilon(y)-y|\le\epsilon$. Define $\mathcal X_\epsilon=q_\epsilon(\mathcal X)\subseteq A_\epsilon(B)$, and let $x_\epsilon$ be the IC rule induced by $\X_\epsilon$ from \cref{prop:finite_ic_characterization}
%

We first bound the allocation error $\Delta_i(d_i):=\abs{x_\epsilon(d_i;B)-x(d_i;B)}.$
Let $m=\inf\mathcal X$ and $\overline x=\sup\mathcal X$, and write the gaps of $\mathcal X$ as \((m,\overline x)\setminus\mathcal X=\bigcup_j(v_j,u_{j+1})\).
On $\X$ we have $x(d_i;B) = d_i$.  Since $\X_\epsilon$ contains an allocation within $\epsilon$ of $d_i$, and $x_\epsilon$ assigns $d_i$ its preferred attainable allocation, the utility definition implies $\Delta_i(d_i) \leq K_i \epsilon \leq A_i \epsilon$.  The same bound holds on the bottom and top plateaus.

Now consider a gap $(v_j, u_{j+1})$ and let $\ell_j = u_{j+1} - v_j$ and $\tau_j = \frac{v_j + \alpha_i u_{j+1}}{1 + \alpha_i}$ be its cutoff via \cref{thm:IC-structure}. Set
\[
I_j := \begin{cases}
    (\tau_j - 2\epsilon, \tau_j + 2\epsilon) \cap (v_j, u_{j+1}) & \ell_j > 2 \epsilon \\
    (v_j, u_{j+1}) & \ell_j \leq 2 \epsilon.
\end{cases}
\]
Thus $|I_j| \leq 4 \epsilon$.  Outside of $I_j$, rounding does not change the selected side of the gap allocated by $x_\epsilon$, so $\Delta_i(d_i) \leq K_i\epsilon$.  On $I_j$ we use the crude bound $\Delta_i(d_i) \leq \ell_j + K_i \epsilon$.  Hence, up to the cutoff points which have $F_i$-measure zero, we have
\[
\E[\Delta_i(d_i)] \leq K_i \epsilon + \sum_{j} \ell_j \Pr(d_i \in I_j) \leq K_i \epsilon + 4 \bar f \epsilon \sum_j \ell_j \leq (K_i + 4 \bar f B) \epsilon \leq A_i \epsilon.
\]

As in the proof of \cref{lem:grid_lipschitz}, the previous bound on $\E[\Delta_i(d_i)]$ gives:
\[
\E[\abs{\beta(x_\epsilon(d_i;B),d_i)-\beta(x(d_i;B),d_i)}] \leq (\bar{f} + A_i)\sqrt{\epsilon}
\]
Hence by the assumed regularity of $W$ and Jensen's inequality,
\[
\E[\abs{W(B-x_\epsilon(d_i;B))-W(B-x(d_i;B))}] \leq L\sqrt{A_i\epsilon}.
\]
Thus the value of $x_\epsilon$ is at least $(\mathcal T_iW)(B)-\eta-\left(\bar f+A_i+L\sqrt{A_i}\right)\sqrt{\epsilon}$.  Letting $\eta \downarrow 0$ shows the lemma
\end{proof}

With the previous two lemmas we are ready to show \cref{thm:grid_approx}.
\begin{proof}[Proof of \cref{thm:grid_approx}]
Let all constants be defined as in \cref{lem:grid_lipschitz,lem:grid_operator_approx} and the statement of \cref{thm:grid_approx}, and define $C_i = \bar f+A_i+L_{i+1}\sqrt{A_i}$.  By \cref{lem:grid_lipschitz}, $V_{i+1}$ is $1/2$-H\"older continuous with constant $L_{i+1}$.  Hence, \cref{lem:grid_operator_approx} gives that for every $B \in G_\epsilon$:
\[
0 \leq (\mathcal T_iV_{i+1})(B)-(\mathcal T_i^\epsilon V_{i+1})(B)\le C_i\sqrt{\epsilon}.
\]
Moreover, $\mathcal T_i^\epsilon$ is nonexpansive in the sup norm, in that for any $W$ and $\tilde W$ defined on $G_\epsilon$:
\[
\sup_{B\in G_\epsilon}
\left|
(\mathcal T_i^\epsilon W)(B)
-
(\mathcal T_i^\epsilon \widetilde W)(B)
\right|
\le
\sup_{B'\in G_\epsilon}|W(B')-\widetilde W(B')|.
\]
Lastly, since $\IC_\epsilon(B;\alpha_i)\subseteq\IC(B;\alpha_i)$, backwards induction also gives that $V_i^\epsilon(B)\le V_i(B)$ for every stage $i$ and grid budget $B$.

Define \(e_i=\sup_{B\in G_\epsilon}\left(V_i(B)-V_i^\epsilon(B)\right)\) and $e_{n+1}=0$.
The one-stage error bound and the nonexpansiveness of $\mathcal T_i^\epsilon$ give \(e_i\le C_i\sqrt{\epsilon}+e_{i+1}\). Iterating backward from $e_{n+1}=0$ gives \(e_1\le \left(\sum_{i=1}^n C_i\right)\sqrt{\epsilon}\). Since $A_i \leq A$ and the recursion for $L_i$ gives $L_{i+1}=r\frac{(q^{n-i}-1)}{(q-1)}$ we have 

\[
\sum_{i=1}^n C_i\le n(\bar f+A)+\sqrt A\sum_{i=1}^nL_{i+1}=n(\bar f+A)+\frac{r\sqrt A}{q-1}\left(\frac{q^n-1}{q-1}-n\right).
\]
Since $B_1\in G_\epsilon$, the bound on $e_1$ and $V_1^\epsilon(B_1)\le V_1(B_1)$ prove the claimed inequality.
\end{proof}

%% file: figures/ic_rule_discrete.tex
\begin{tikzpicture}[
    x=1.25cm,
    y=1.00cm,
    every node/.style={
        paper figure text,
        font=\normalsize
    },
    axis/.style={
        paper axis,
        line width=1.2pt
    },
    gridline/.style={
        paper grid,
        line width=0.55pt
    },
    diag/.style={
        paper benchmark,
        line width=1.4pt
    },
    rule/.style={
        paper primary,
        line width=2.0pt
    },
    pointset/.style={
        royalblue,
        fill=royalblue
    },
    gridpt/.style={
        black!28,
        fill=black!28
    },
    annotation/.style={
        <->,
        paper auxiliary,
        line width=1.0pt
    },
    arrowlabel/.style={
        paper tick label,
        font=\normalsize
    },
]

\def\B{4.0}

\def\akone{0.75}
\def\aktwo{1.65}
\def\akthree{2.70}
\def\akfour{3.55}

\def\tone{1.29}
\def\ttwo{2.28}
\def\tthree{3.21}

\def\leftx{0}
\def\rightx{2.35}

\def\xstretch{1.85}
\def\gap{0.04}

\begin{scope}[shift={(\leftx,0)}]

    \draw[axis] (0,0) -- (0,\B);

    \node[left] at (0,0) {$0$};
    \node[left] at (0,\B) {$B$};

    \draw[
        paper auxiliary,
        line width=1.0pt
    ]
        (0.9,0) -- (0.9,\B);

    \foreach \yy in {
        0.35,0.75,1.10,1.65,2.10,2.70,3.10,3.55
    }{
        \filldraw[gridpt]
            (0.9,\yy) circle (1.2pt);
    }

    \foreach \yy/\lab in {
        \akone/$a_{k_1}$,
        \aktwo/$a_{k_2}$,
        \akthree/$a_{k_3}$,
        \akfour/$a_{k_4}$
    }{
        \filldraw[pointset]
            (0.9,\yy) circle (1.8pt);

        \draw[axis, line width=1.2pt]
            (-0.055,\yy) -- (0.055,\yy);

        \node[left]
            at (-0.055,\yy)
            {\lab};
    }

    \node[below]
        at (0.9,-0.35)
        {attainable set $\mathcal X\subseteq A_\epsilon(B)$};

\end{scope}

\begin{scope}[
    shift={(\rightx,0)},
    xscale=\xstretch
]

    \draw[axis] (0,0) -- (0,\B);
    \draw[axis] (0,0) -- (\B,0);

    \node[left] at (0,0) {$0$};
    \node[left] at (0,\B) {$B$};

    \node[below]
        at (\B/2,-0.43)
        {$d$};

    \node[rotate=90]
        at (-0.33,\B/2)
        {$x(d;B)$};

    \foreach \xx in {
        \tone,\ttwo,\tthree
    }{
        \draw[gridline]
            (\xx,0) -- (\xx,\B);
    }

    \foreach \yy in {
        \akone,\aktwo,\akthree,\akfour
    }{
        \draw[gridline]
            (0,\yy) -- (\B,\yy);
    }

    \foreach \xx/\lab in {
        \tone/$t_1$,
        \ttwo/$t_2$,
        \tthree/$t_3$
    }{
        \node[below]
            at (\xx,0)
            {\lab};
    }

    \foreach \yy/\lab in {
        \akone/$a_{k_1}$,
        \aktwo/$a_{k_2}$,
        \akthree/$a_{k_3}$,
        \akfour/$a_{k_4}$
    }{
        \node[left]
            at (0,\yy)
            {\lab};
    }

    \draw[diag]
        (0,0) -- (\B,\B);

    \draw[rule]
        (0,\akone) -- (\tone,\akone);

    \draw[rule]
        (\tone,\akone) -- (\tone,\aktwo);

    \draw[rule]
        (\tone,\aktwo) -- (\ttwo,\aktwo);

    \draw[rule]
        (\ttwo,\aktwo) -- (\ttwo,\akthree);

    \draw[rule]
        (\ttwo,\akthree) -- (\tthree,\akthree);

    \draw[rule]
        (\tthree,\akthree) -- (\tthree,\akfour);

    \draw[rule]
        (\tthree,\akfour) -- (\B,\akfour);

    \draw[diag]
        (0.12,3.72) -- (0.42,3.72);

    \node[anchor=west]
        at (0.50,3.72)
        {$x(d;B)=d$};

    \draw[rule]
        (0.12,3.38) -- (0.42,3.38);

    \node[anchor=west]
        at (0.50,3.38)
        {induced IC rule};

\end{scope}

\end{tikzpicture}

%% file: figures/dag_diagram.tex
\begin{tikzpicture}[
    node distance=2.2cm and 3.0cm,
    every node/.style={paper figure text},
    state/.style={circle, paper axis, fill=white, minimum size=8mm, inner sep=1pt},
    pair/.style={rectangle, paper axis, fill=white, rounded corners=1.5pt, minimum height=8mm, minimum width=18mm, inner sep=2pt},
    edge/.style={-{Stealth[length=2mm,width=1.35mm]}, paper axis},
    lab/.style={paper figure text, fill=white, inner sep=1pt}
]


\node[state] (s) {$s$};
\node[pair, right=of s, yshift=1.0cm] (pairAAp) {$(a,a')$};
\node[pair, right=of s, yshift=-1.0cm] (pairApApp) {$(a',a'')$};
\node[state, right=4.0cm of pairAAp, yshift=-1.0cm] (t) {$t$};

\draw[edge] (s) -- 
    node[lab, above] {$\Gamma_{i,b}(a,0,\tau(a,a'))$} 
    (pairAAp);

\draw[edge] (s) -- 
    node[lab, below] {$\Gamma_{i,b}(a',0,\tau(a',a''))$} 
    (pairApApp);


\draw[edge] (pairAAp) -- 
    node[lab, right] {$\Gamma_{i,b}(a',\tau(a,a'),\tau(a',a''))$} 
    (pairApApp);


\draw[edge] (pairAAp) -- 
    node[lab, above] {$\Gamma_{i,b}(a',\tau(a,a'),1)$} 
    (t);

\draw[edge] (pairApApp) -- 
    node[lab, below] {$\Gamma_{i,b}(a'',\tau(a',a''),1)$} 
    (t);

\end{tikzpicture}

%% file: short_parts/simulations.tex
\section{Numerical Experiment}
\label{sec:numerics}

The preceding analysis characterizes optimal incentive-compatible mechanisms, but does not highlight the importance of accounting for strategic reporting. In this section we use numerical simulations to measure the consequences of ignoring agents' reporting incentives and the gains from explicitly accounting for them.  In particular, we ask:
{\em
(i) How much does the planner lose when implementing an allocation rule that assumes truthful reporting when agents report their demands strategically?
(ii) How much of this loss can optimal incentive-compatible design recover?
} We address these questions by comparing the following benchmarks (see \cref{ec:numerical-details} for additional implementation details):
\begin{itemize}
    \item {\bf Optimal IC.} Our first benchmark is the optimal IC value $V_1(B_1)$, defined in \cref{sec:prelim}, which corresponds to the planner's objective value under the optimal incentive-compatible mechanism.
    \item {\bf First-best.}  We next consider the case in which the
planner chooses an optimal allocation rule under the assumption
of truthful reporting, and agents report their demands truthfully.
We refer to this as the ``first-best'' allocation rule.  The first-best value is
    \(
    V_1^{FB} := \sup_{\bfx} V^{\bfx}(B_1),
    \)
    and we let $\bfx^{FB}$ be the corresponding optimal allocation rule. Note that the difference between this rule and the optimal IC mechanism is that we do not enforce the IC constraint.
    \item {\bf Naive first-best.} We next consider what happens when the planner announces the first-best rule, but agents misreport their demands strategically.  For each agent $i$ with true demand $d_i$ and a remaining budget $B_i$, let
    \[
    \hat{d}_i^{FB}(d_i; B_i) \in \argmax_{d \in [0,1]} u(x_i^{FB}(d, B_i), d_i;\alpha_i)
    \]
    denote the reported demand for agent $i$ when the planner proposes $x_i^{FB}$. We define
    \(
    x_i^N(d_i;B_i) := x_i^{FB}(\hat{d}_i^{FB}; B_i).
    \)
    This represents the allocation rule induced by announcing $\bfx^{FB}$ to strategic agents, which achieves a value of $V_1^N := V^{\bfx^N}(B_1)$. 
\end{itemize}
We compare the performance through two different metrics:
\begin{itemize}
    \item {\bf Cost of ignoring strategic reporting}: First, we define $\Delta^N := V_1^{FB} - V_1^N$ which represents the loss due to naively implementing the first-best rule when agents report strategically.
    \item {\bf Value of incentive-compatible design}: Second, we consider $\Delta^G := V_1(B_1) - V_1^N$ that measures how much of this loss the planner can recover by using the optimal IC mechanism instead of the naive first-best implementation.
\end{itemize}

\paragraph{Experimental design.}
We consider $n=3$ agents with independent demands drawn from a common distribution $F$ and a common overage penalty $\alpha\in\{0.1,1,5\}$. 
In all settings, we discretize remaining budgets and allocations using grid spacing $\epsilon=10^{-4}$, rounding initial budgets to this grid.
We consider three families of demand distributions, each characterized by its mean $\mu$ and coefficient of variation $CV$. In each family, a positive uniform mixture component ensures full support over $[0,1]$:
\begin{itemize}
    \item {\em Uniform-Window}:
\(
    F^{UW}_{\mu,CV}
    =
    \rho_{\mu,CV} U[0,1]
    +
    (1-\rho_{\mu,CV})U[a_{\mu,CV},b_{\mu,CV}].
\)
    \item {\em Single-Peaked}:
\(
    F^{SP}_{\mu,CV}
    =
    \rho_{\mu,CV} U[0,1]
    +
    (1-\rho_{\mu,CV})\operatorname{Beta}(p_{\mu,CV},q_{\mu,CV}).
\)
    \item {\em Double-Peaked}: This considers a mixture of two Beta distributions via
\[
    F^{DP}_{\mu,CV}
    =
    \rho_{\mu,CV} U[0,1]
    +
    (1-\rho_{\mu,CV})
    \left[
        \lambda_{\mu,CV}\operatorname{Beta}(p^1_{\mu,CV},q^1_{\mu,CV})
        +
        (1-\lambda_{\mu,CV})\operatorname{Beta}(p^2_{\mu,CV},q^2_{\mu,CV})
    \right].
\]
\end{itemize}
\noindent For each distribution family, we calibrate the parameters to obtain mean $\mu \in \{0.05,0.10,0.15,0.20\}$ and coefficient of variation $CV \in \{0.50,0.75,1.00,1.50\}$. The full parameter specifications are reported in \ref{ec:numerical-details}.
For the initial budget $B_1$ we set $B_1 = t \sum_{i=1}^n \mathbb{E}[d_i]$ where $t$ is the so-called ``tightness level''.  We test values of $t$ in $\{0.2, 0.4, \ldots, 2.0\}$, where a larger $t$ corresponds to more abundant inventory. Demands are integrated using midpoint quadrature on \(100\) equally spaced bins on \([0,1]\). This setting results in a total of 1440 scenarios that we examine in our numerical study.

\paragraph{Numerical findings.} We exam how these metrics vary with resource availability, the overage penalty, and demand variability.

\begin{observation}[Cost of Ignoring Strategic Reporting]
    The cost of ignoring strategic reporting is largest at intermediate tightness levels.
\end{observation}

\begin{figure}[!t]
    \centering
    \begin{subfigure}[t]{0.49\textwidth}
        \centering
        \input{figures/strategic_main_gaps_by_tightness_quantile}
    \caption{$\Delta^N$ and $\Delta^G$ across tightness levels.}        \label{fig:gaps-tightness}
    \end{subfigure}
    \hfill
    \begin{subfigure}[t]{0.49\textwidth}
        \centering
        \input{figures/DeltaGalphaCV.tex}
        \caption{$\Delta^G$ across overage penalties, by $CV$.}
        \label{fig:gain-alpha}
    \end{subfigure}
    \caption{Average gaps over the remaining parameters.  Shading in panel~(a) shows the 5th--95th percentiles.}    \label{fig:strategic-main-gaps}
\end{figure}
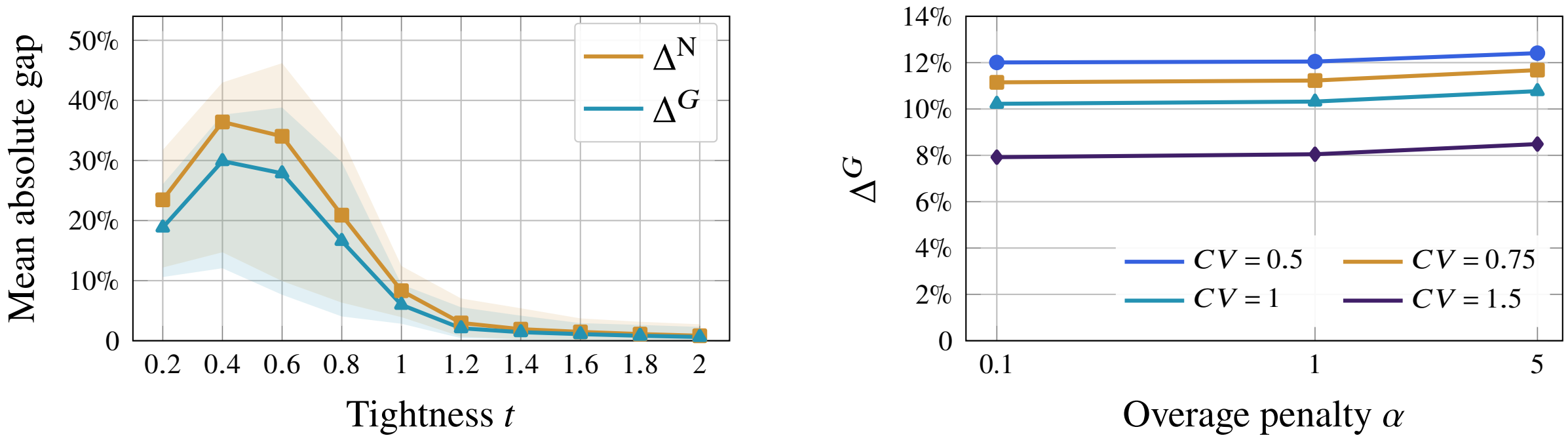

In \cref{fig:strategic-main-gaps}(\subref{fig:gaps-tightness}) we observe that $\Delta^N$, the cost of ignoring strategic reporting, is largest at intermediate tightness levels. Averaging over all parameters, the gap is $36.4\%$ at $t=0.4$ and $34.0\%$ at $t=0.6$. This pattern reflects how the first-best allocation rule responds to reported demand. When resources are scarce, the first-best rule limits allocations even for high reported demands, restricting the gains agents can obtain by exaggerating their needs. On the other hand, when resources are abundant, the first-best rule allocates agents their full demand leaving not much to gain by misreporting. At intermediate resource levels, agents have the greatest scope to obtain more than the first-best rule would allocate under truthful reporting, making it especially important for the planner to account for strategic reporting.

\begin{observation}[Value of Incentive-Compatible Design]
If agents are strategic, the planner benefits more, on average, from implementing an incentive-compatible mechanism when the overage penalty $\alpha$ is higher and demand variability is lower.
\end{observation}

In \cref{fig:strategic-main-gaps}(\subref{fig:gain-alpha}) we observe that $\Delta^G$, the value of incentive-compatible design, decreases with demand variability and increase with overage penalty. For instance, at $\alpha=0.1$, increasing $CV$ from $0.5$ to $1.5$ reduces the average gain from $12.01\%$ to $7.92\%$; at $CV=1.5$, increasing $\alpha$ from $0.1$ to $5$ raises it from $7.92\%$ to $8.49\%$.
The planner enforces truthful reporting by allocating more than requested at some demand levels, so agents with lower needs would incur an overage cost by overreporting their demand. This over-allocation leaves fewer resources for later agents, potentially reducing the planner's objective value. With a higher overage penalty, the planner can induce truthful reporting while limiting over-allocation to a narrower range of reported demands. Lower demand variability can also reduce the need to preserve resources for potentially very high future demands, making over-allocation less costly for the planner. The gain is therefore high when the overage penalty is large and demand variability is low.

%% file: figures/strategic_main_gaps_by_tightness_quantile.tex
\begin{tikzpicture}
\begin{axis}[
  paper plot,
  xlabel style={font=\normalsize},
 ylabel style={font=\normalsize},
  ylabel={Mean absolute gap},
  ymin=0,
  enlarge y limits={upper,value=0.08},
  ymax=0.50,
  ytick={0,0.1,0.2,0.3,0.4,0.5},
  yticklabels={0,10\%,20\%,30\%,40\%,50\%},
  legend style={at={(0.98,0.98)},anchor=north east,  font=\normalsize},
  legend image post style={mark=none}
]
\addplot[draw=none,forget plot,name path=lower2] coordinates {
  (0.2,0.12179515) (0.4,0.147194396) (0.6,0.0992997847) (0.8,0.0633966116) (1,0.0392419774) (1.2,0.00685151438) (1.4,0.00225158839) (1.6,0.00160477402) (1.8,0.00154755997) (2,0.00141539738)
};
\addplot[draw=none,forget plot,name path=upper2] coordinates {
  (0.2,0.317533896) (0.4,0.429838954) (0.6,0.461942578) (0.8,0.337369854) (1,0.124409713) (1.2,0.0703624469) (1.4,0.0540616028) (1.6,0.0371968358) (1.8,0.0315771402) (2,0.0274297606)
};
\addplot[paper band two,forget plot] fill between[of=lower2 and upper2];
\addplot[draw=none,forget plot,name path=lower3] coordinates {
  (0.2,0.105982541) (0.4,0.120646329) (0.6,0.076523469) (0.8,0.0403904294) (1,0.0282357659) (1.2,0.00423702016) (1.4,0.001792245) (1.6,0.00134585446) (1.8,0.00134696247) (2,0.000972520225)
};
\addplot[draw=none,forget plot,name path=upper3] coordinates {
  (0.2,0.260511115) (0.4,0.376107018) (0.6,0.388149787) (0.8,0.297294846) (1,0.0937042073) (1.2,0.0556733451) (1.4,0.0417496642) (1.6,0.0293198054) (1.8,0.0263145636) (2,0.0229779593)
};
\addplot[paper band three,forget plot] fill between[of=lower3 and upper3];
\addplot[paper series two] coordinates {
  (0.2,0.234527735) (0.4,0.364122816) (0.6,0.340275461) (0.8,0.209024126) (1,0.08319679) (1.2,0.0295712604) (1.4,0.0192456569) (1.6,0.0146884966) (1.8,0.0110797967) (2,0.00828567574)
};
\addlegendentry{$\DeltaN$}   
\addplot[paper series three] coordinates {
  (0.2,0.188389016) (0.4,0.298955396) (0.6,0.278665126) (0.8,0.166088644) (1,0.0598019991) (1.2,0.0208039137) (1.4,0.0142378497) (1.6,0.0110032552) (1.8,0.00827720004) (2,0.00617169505)
};
\addlegendentry{$\Delta^G$}  
\end{axis}
\end{tikzpicture}

%% file: figures/DeltaGalphaCV.tex
\begin{tikzpicture}
\begin{semilogxaxis}[
  paper plot,
  xlabel style={font=\normalsize},
  ylabel style={font=\normalsize},
  xlabel={Overage penalty $\alpha$},
  ylabel={$\Delta^{G}$},
  xmin=0.08,
  xmax=6,
  xtick={0.1,1,5},
  xticklabels={0.1,1,5},
  ymin=0,
  ymax=0.14,
  ytick={0,0.02,0.04,0.06,0.08,0.10,0.12,0.14},
  yticklabels={0,2\%,4\%,6\%,8\%,10\%,12\%,14\%},
  enlarge x limits=false,
  enlarge y limits=false,
  legend columns=2,
  legend style={
    at={(0.98,0.04)},
    anchor=south east,
    draw=none,
    /tikz/every even column/.append style={column sep=0.8em}
  },
  legend image post style={mark=none}
]

\addplot[paper series one] coordinates {
  (0.1,0.1200794421)
  (1,0.1204731360)
  (5,0.1240976204)
};
\addlegendentry{$CV=0.5$}

\addplot[paper series two] coordinates {
  (0.1,0.1114952714)
  (1,0.1122761403)
  (5,0.1167765505)
};
\addlegendentry{$CV=0.75$}

\addplot[paper series three] coordinates {
  (0.1,0.1022382156)
  (1,0.1031985168)
  (5,0.1077093530)
};
\addlegendentry{$CV=1$}

\addplot[
  paper series one,
  color=nupurple,
  mark=diamond*
] coordinates {
  (0.1,0.0792055123)
  (1,0.0804666342)
  (5,0.0848565214)
};
\addlegendentry{$CV=1.5$}

\end{semilogxaxis}
\end{tikzpicture}

%% file: short_parts/conclusion.tex
\section{Conclusion}
\label{sec:conclusion}

Private information creates a central challenge in allocating scarce resources since agents may overstate their needs to obtain larger allocations. This paper studies incentive-compatible mechanisms for sequential resource allocation when fairness is the primary objective and monetary transfers are unavailable or inappropriate. These concerns are particularly relevant to humanitarian operations and disaster relief, where decision-makers must distribute limited supplies while relying on recipients' reports of their needs.
We characterize deterministic incentive-compatible mechanisms through their closed attainable sets and identify conditions under which cap mechanisms are optimal. Beyond these cases, we develop an approximation framework that discretizes the allocation space and solves the resulting problem through a sequence of longest-path problems on directed acyclic graphs.

Our numerical results show that ignoring strategic reporting can lead to substantial losses, particularly at intermediate resource levels. Optimal incentive-compatible design recovers part of this loss, with the largest gains occurring when demand variability is low and the overage penalty is high. Planners should therefore pay particular attention to strategic behavior when resource availability is intermediate, demand variability is low, and overage penalties are high.
Two directions merit further investigation. The first is to allow the planner to choose the order in which agents are visited, connecting incentive-compatible allocation with routing and sequencing decisions. The second is to consider populations containing both truthful and strategic agents. This would connect the truthful-reporting benchmark with the fully strategic setting studied here and clarify how allocation mechanisms should account for differences in agents' reporting behavior.

%% file: short_parts/app_simulations.tex
\section{Numerical Experiment Details}
\label{ec:numerical-details}

The parameters used to calibrate the three distribution families are reported in \cref{tab:dist_calibrations}.  Since demands are normalized to lie in \([0,1]\), individual allocations are restricted to \([0,1]\). The total budget \(B_1\), however, may exceed one when tightness is high. In such cases the budget grid is extended beyond one, while each individual allocation remains bounded above by one.

\begin{table}[!h]
\centering
\caption{Demand distribution calibration parameters. Each subtable reports the calibrated parameters for one demand family as a function of the target mean $\mu$ and coefficient of variation $CV$.  See \cref{sec:numerics} for the CDFs of each distribution family.}
\label{tab:dist_calibrations}

\begin{subtable}{\textwidth}
\centering
\caption{Uniform-window family. Each cell reports $(\rho,a,b)$.}
\label{tab:dist_uniform_window}
\begin{adjustbox}{max width=\textwidth}
\begin{tabular}{ccccc}
\toprule
\(\mu \backslash CV\) & 0.50 & 0.75 & 1.00 & 1.50 \\
\midrule
0.05 &
\((0.001092,\,0.018866,\,0.080150)\) &
\((0.003510,\,0.013693,\,0.083137)\) &
\((0.007414,\,0.013215,\,0.080062)\) &
\((0.018422,\,0.011816,\,0.071293)\) \\
0.10 &
\((0.005102,\,0.036449,\,0.159449)\) &
\((0.016698,\,0.025188,\,0.161227)\) &
\((0.035151,\,0.023292,\,0.147563)\) &
\((0.084725,\,0.017577,\,0.108369)\) \\
0.15 &
\((0.013443,\,0.052373,\,0.238088)\) &
\((0.044934,\,0.034087,\,0.232979)\) &
\((0.093238,\,0.029984,\,0.198038)\) &
\((0.210443,\,0.015928,\,0.097499)\) \\
0.20 &
\((0.027984,\,0.066186,\,0.316540)\) &
\((0.096053,\,0.040181,\,0.296063)\) &
\((0.192764,\,0.033042,\,0.223681)\) &
\((0.389852,\,0.002392,\,0.014240)\) \\
\bottomrule
\end{tabular}
\end{adjustbox}
\end{subtable}

\medskip

\begin{subtable}{\textwidth}
\centering
\caption{Single-peaked beta family. Each cell reports $(\rho,p,q)$.}
\label{tab:dist_single_peaked_beta}
\begin{adjustbox}{max width=\textwidth}
\begin{tabular}{ccccc}
\toprule
\(\mu \backslash CV\) & 0.50 & 0.75 & 1.00 & 1.50 \\
\midrule
0.05 &
\((0.001092,\,7.394341,\,141.961957)\) &
\((0.003510,\,5.501774,\,108.136181)\) &
\((0.007414,\,5.522234,\,112.882111)\) &
\((0.018422,\,5.572557,\,128.529755)\) \\
0.10 &
\((0.005102,\,6.766357,\,62.314241)\) &
\((0.016698,\,5.014984,\,48.789600)\) &
\((0.035151,\,5.100869,\,54.609149)\) &
\((0.084725,\,5.346361,\,79.553064)\) \\
0.15 &
\((0.013443,\,6.127439,\,36.063593)\) &
\((0.044934,\,4.553294,\,29.545300)\) &
\((0.093238,\,4.779338,\,37.140570)\) &
\((0.210443,\,5.415124,\,90.067066)\) \\
0.20 &
\((0.027984,\,5.478102,\,23.148638)\) &
\((0.096053,\,4.141206,\,20.490921)\) &
\((0.192764,\,4.613704,\,31.329353)\) &
\((0.389852,\,5.854014,\,698.090561)\) \\
\bottomrule
\end{tabular}
\end{adjustbox}
\end{subtable}

\medskip

\begin{subtable}{\textwidth}
\centering
\caption{Double-peaked beta-mixture family. Each cell reports
$(\rho,\lambda,p_1,q_1,p_2,q_2)$.}
\label{tab:dist_double_peaked_beta}
\scriptsize
\setlength{\tabcolsep}{2.5pt}
\renewcommand{\arraystretch}{1.05}
\resizebox{\textwidth}{!}{%
\begin{tabular}{ccccc}
\toprule
\(\mu \backslash CV\) & 0.50 & 0.75 & 1.00 & 1.50 \\
\midrule
0.05 &
\makecell[c]{\((0.001092,\,0.145090,\,9.901613,\)\\\(990.098387,\,56.229846,\,943.770154)\)} &
\makecell[c]{\((0.003510,\,0.191886,\,9.682987,\)\\\(990.317013,\,57.611813,\,942.388187)\)} &
\makecell[c]{\((0.007414,\,0.190820,\,9.327756,\)\\\(990.672244,\,55.437445,\,944.562555)\)} &
\makecell[c]{\((0.018422,\,0.187639,\,8.310903,\)\\\(991.689097,\,49.233103,\,950.766897)\)} \\
0.10 &
\makecell[c]{\((0.005102,\,0.160473,\,19.589747,\)\\\(980.410253,\,112.926793,\,887.073207)\)} &
\makecell[c]{\((0.016698,\,0.207888,\,18.641475,\)\\\(981.358525,\,112.776977,\,887.223023)\)} &
\makecell[c]{\((0.035151,\,0.205773,\,17.085471,\)\\\(982.914529,\,103.133732,\,896.866268)\)} &
\makecell[c]{\((0.084725,\,0.198515,\,12.594575,\)\\\(987.405425,\,75.450755,\,924.549245)\)} \\
0.15 &
\makecell[c]{\((0.013443,\,0.169392,\,29.046168,\)\\\(970.953832,\,168.925154,\,831.074846)\)} &
\makecell[c]{\((0.044934,\,0.218148,\,26.706636,\)\\\(973.293364,\,163.339271,\,836.660729)\)} &
\makecell[c]{\((0.093238,\,0.213247,\,22.802235,\)\\\(977.197765,\,138.733042,\,861.266958)\)} &
\makecell[c]{\((0.210443,\,0.195908,\,11.342690,\)\\\(988.657310,\,67.767554,\,932.232446)\)} \\
0.20 &
\makecell[c]{\((0.027984,\,0.177956,\,38.272621,\)\\\(961.727379,\,224.504123,\,775.495877)\)} &
\makecell[c]{\((0.096053,\,0.227320,\,33.624427,\)\\\(966.375573,\,207.690981,\,792.309019)\)} &
\makecell[c]{\((0.192764,\,0.216841,\,25.672294,\)\\\(974.327706,\,156.794052,\,843.205948)\)} &
\makecell[c]{\((0.389852,\,0.072573,\,1.663203,\)\\\(998.336797,\,8.836612,\,991.163388)\)} \\
\bottomrule
\end{tabular}%
}
\end{subtable}
\end{table}

\paragraph{Tie-breaking and discretization effects.} 
We first specify how ties are resolved in our numerical implementation. When several allocations maximize the first-best objective, we select the smallest allocation.  Moreover, when an agent is indifferent among utility-maximizing reports, we break ties in the planner's favor. The first convention selects a specific first-best rule while avoiding unnecessary resource use. The second specifies agents' behavior when several reports are equally attractive. Under our continuous demand distributions, indifference between distinct attainable allocations occurs only at demand cutoffs with probability zero, while ties between reports yielding the same allocation are immaterial. Consequently, changing agents' tie-breaking does not affect expected performance or the resulting numerical insights.

We next note a small discretization effect associated with the overage penalties $\alpha$.  In the continuous problem, both $V_1^{FB}$ and $V_1^N$ are independent of $\alpha$.  
Indeed, the first-best value is independent of $\alpha$ because neither the planner's objective nor its constraints depend on the agents' utility.  Moreover, under the first-best rule, every allocation between zero and the maximum attainable allocation can be induced by some reported demand. A strategic agent therefore obtains their true demand whenever it is attainable, and otherwise the largest attainable allocation, regardless of $\alpha$.

On the finite allocation grid used in our computations, however, an agent's true demand might not be attainable. The preferred allocation can then depend on $\alpha$. For adjacent attainable allocations $a < a'$, the agent switches from $a$ to $a'$ at the cutoff $\frac{a + \alpha a'}{1 + \alpha}$.  Thus, the computed value $V_1^N$ can vary slightly with $\alpha$ even though its continuous counterpart does not. 
However, at $\epsilon=10^{-4}$, the range of the computed naive value across $\alpha\in\{0.1,1,5\}$, holding all other parameters fixed, averages $6.95\times10^{-4}$ and reaches at most $8.71\times10^{-3}$ across the 480 parameter combinations. The computed first-best value is identical across all three values of $\alpha$.